\documentclass[sigconf, screen]{acmart}

\usepackage[most]{tcolorbox}
\usepackage{caption}
\usepackage{booktabs}
\usepackage{xspace}
\usepackage{enumitem}
\usepackage[flushleft]{threeparttable}
\usepackage{makecell}
\usepackage{multirow}
\usepackage{xcolor, colortbl}
\usepackage{subcaption}
\usepackage{tikz}
\usepackage{wrapfig}
\usepackage{textcomp}
\usepackage{scalerel}
\usepackage{tipa}
\usepackage{pifont}
\usepackage[subtle]{savetrees}
\usepackage[most]{tcolorbox}
\usepackage{listings}
\usepackage{caption}
\usepackage[normalem]{ulem}
\usepackage[utf8]{inputenc}
\usepackage{lipsum}
\usepackage{lettrine}
\usepackage{amsmath}
\usepackage{amsfonts}
\usepackage{graphicx}
\usepackage{array}
\usepackage{tabularx}
\usepackage{bbding}
\usepackage{makecell}
\usepackage{caption}
\usepackage{subcaption}
\usepackage{enumitem}
\usepackage{booktabs}
\usepackage{longtable}
\usepackage[table]{xcolor}
\usepackage{geometry}
\usepackage{scalefnt}

\definecolor{bg_human}{RGB}{240, 242, 245}
\definecolor{bg_7b}{RGB}{220, 235, 250}
\definecolor{bg_14b}{RGB}{225, 242, 235}
\definecolor{bg_32b}{RGB}{250, 240, 225}
\definecolor{bg_ours}{RGB}{255, 235, 235}
\definecolor{small_models}{RGB}{220, 235, 250}
\definecolor{large_models}{RGB}{225, 242, 235}	
\definecolor{log_models}{RGB}{250, 240, 225}

\definecolor{pos}{RGB}{226,226,226}
\newcommand{\highlightcolor}{pos}

\newtcolorbox{quotebox}{
  enhanced,
  colback=white,
  colframe=black!45,
  boxrule=0.6pt,
  arc=2.5mm,
  left=4mm, right=4mm, top=2mm, bottom=2mm,
  boxsep=0pt
}

\AtBeginDocument{%
  }

\copyrightyear{2026}
\acmYear{2026}
\setcopyright{cc}
\setcctype{by-nc-nd}
\acmConference[ASE '26]{Proceedings of the 41st IEEE/ACM International Conference on Automated Software Engineering}{October 12--16, 2026}{Munich, Germany}
\acmBooktitle{Proceedings of the 41st IEEE/ACM International Conference on Automated Software Engineering (ASE '26), October 12--16, 2026, Munich, Germany}
\acmDOI{10.1145/3832783.3834433}
\acmISBN{979-8-4007-2882-2/2026/10}

\begin{document}

\newcommand{\icon}{\scalerel*{\includegraphics{figs/Bifrost.png}}{\textrm{C}}\xspace}

\title{\textsc{Bifrost}: Empowering Pretrained Language Model with Fallibility Representation for Log-Based Fault Diagnosis}

\settopmatter{authorsperrow=4}

\author{Minghua He}
\affiliation{%
  \institution{Peking University, Institute for Artificial Intelligence}
  \city{Beijing}
  \country{China}
}
\email{hemh2120@stu.pku.edu.cn}

\author{Tong Jia}
\authornote{Corresponding Authors}
\affiliation{%
  \institution{Peking University, Institute for Artificial Intelligence}
  \city{Beijing}
  \country{China}
}
\email{jia.tong@pku.edu.cn}

\author{Lingzhe Zhang}
\affiliation{%
  \institution{Peking University, Institute for Artificial Intelligence}
  \city{Beijing}
  \country{China}
}
\email{zhang.lingzhe@stu.pku.edu.cn}

\author{Chiming Duan}
\affiliation{%
  \institution{Peking University, Institute for Artificial Intelligence}
  \city{Beijing}
  \country{China}
}
\email{duanchiming@stu.pku.edu.cn}

\author{Xinlong Zhao}
\affiliation{%
  \institution{Peking University}
  \city{Beijing}
  \country{China}
}
\email{xinlongzhao1126@gmail.com}

\author{Leyi Pan}
\affiliation{%
  \institution{Tsinghua University}
  \city{Beijing}
  \country{China}
}
\email{panly24@mails.tsinghua.edu.cn}

\author{Cheng Wang}
\affiliation{%
  \institution{Alibaba Group}
  \city{Hangzhou}
  \country{China}
}
\email{wc189854@alibaba-inc.com}

\author{Kangjin Wang}
\affiliation{%
  \institution{Alibaba Group}
  \city{Hangzhou}
  \country{China}
}
\email{kangjin.wkj@alibaba-inc.com}

\author{Yinghao Yu}
\affiliation{%
  \institution{Alibaba Group}
  \city{Hangzhou}
  \country{China}
}
\email{yinghao.yyh@alibaba-inc.com}

\author{Liping Zhang}
\affiliation{%
  \institution{Alibaba Group}
  \city{Hangzhou}
  \country{China}
}
\email{liping.z@alibaba-inc.com}

\author{Yifan Wu}
\affiliation{%
  \institution{Peking University, Beijing Key Laboratory of Data Intelligence and Security}
  \city{Beijing}
  \country{China}
}
\email{yifanwu@pku.edu.cn}

\author{Ying Li}
\authornotemark[1]
\affiliation{%
  \institution{Peking University, Beijing Key Laboratory of Data Intelligence and Security}
  \city{Beijing}
  \country{China}
}
\email{li.ying@pku.edu.cn}

\renewcommand{\shortauthors}{Minghua He et al.}

\begin{abstract}
Log-based fault diagnosis is crucial for runtime debugging and maintenance.
Existing fault diagnosis methods use language models pre-trained on natural language (PLMs) for log representation. 
However, system faults are reflected in the multi-level structure of system logs. 
PLMs pre-trained on natural language struggle to comprehensively capture multi-level fault information, failing to meet the requirements of fault diagnosis. 
We refer to this information as fallibility representations. 
To address this problem, we propose a novel log representation learning method, \textsc{Bifrost}. 
It draws inspiration from the log analysis experience of Site Reliability Engineers and meticulously designs strategies based on self-supervised contrastive learning to learn the fallibility representations of logs. 
Across three public systems and one industrial ML-as-a-Service system, the log representations produced by \textsc{Bifrost} outperform existing PLMs by average margins of 9.83\% in F1 for anomaly detection, 18.28\% in HR@k for root cause localization, and 20.88\% in Macro-F1 for fault identification.
\end{abstract}




\maketitle

\section{Introduction}

\begin{center}
{\itshape\bfseries ``You see, but you do not observe.''\par}

\begin{flushright}
{\normalfont\small\color{black!70} --- Arthur Conan Doyle, \textit{A Scandal in Bohemia}}
\end{flushright}
\end{center}

Modern software systems are composed of many interacting components and evolve rapidly, which complicates their execution logic and expands the space of potential faults \cite{niu2023locating, lin2023edits, duan2026aims, zhang2026towards, zhang2026e2e, luo2021ntam, wu2024falcon, liu2021reproducibility}. Even brief outages can incur substantial losses. For example, a one-hour Prime Day disruption at Amazon has been estimated to cost up to 100 million USD \cite{yu2023cmdiagnostor}. To reduce time-to-recovery and improve availability, providers must quickly identify root causes of anomalies.
Fault diagnosis therefore provides guidance for debugging and recovery, improving system reliability \cite{wong2016survey, zhang2024failure, zhang2026runtimeslicer, zhang2026efficient, jiang2011efficient}.

\begin{figure}[t]
\captionsetup{skip=2pt}
\centerline{
\includegraphics[width=0.9\linewidth]{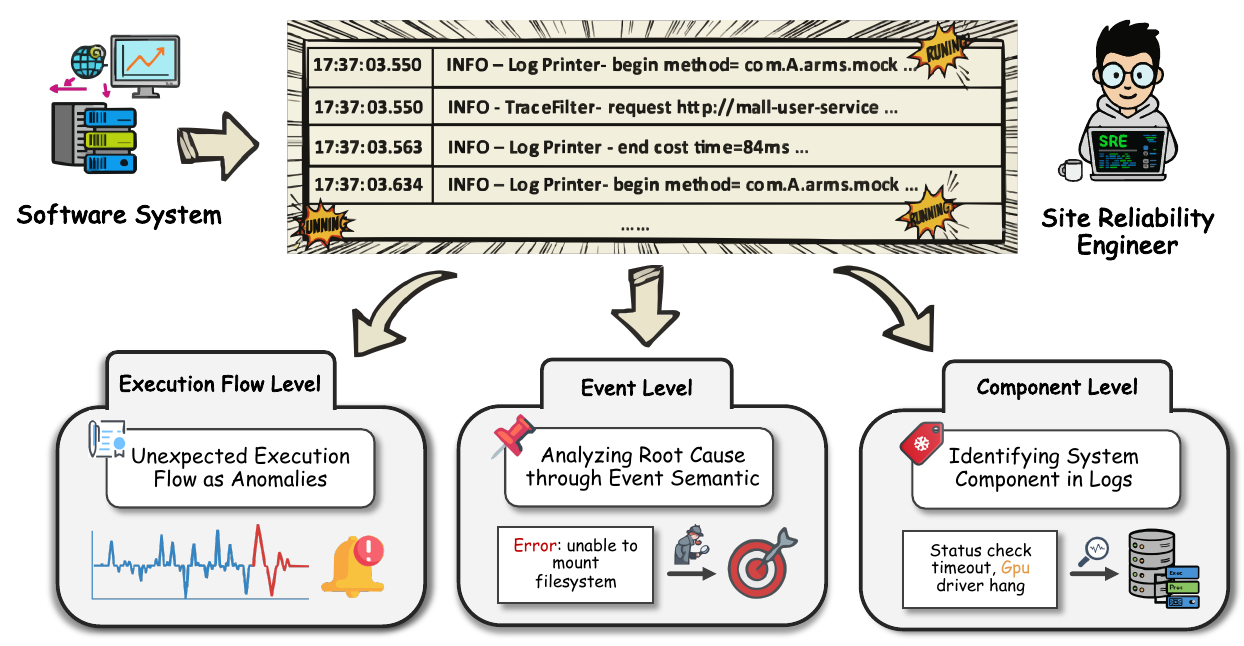}
}
\caption{Practical log analysis experience of SREs.}
\Description{Practical log analysis experience of SREs.}
\label{fig:teaser}
\vspace{-0.4cm}
\end{figure}

Logs record system operation and are vital for fault diagnosis, which has evolved along two primary paradigms. 
One paradigm uses Large Language Models (LLMs) to directly analyze raw logs or logs retrieved from external memory for semantic reasoning, interactive troubleshooting, and tool-augmented diagnosis \cite{zhang2025survey, jiang2025l4,huang2025no, xiao2025coorlog}. 
The other paradigm follows the more established workflow of log parsing, log representation, and downstream diagnosis \cite{zhang2019robust,du2017deeplog,meng2019loganomaly}: semi-structured logs are first parsed into structured log events \cite{he2017drain, sun2025exploring}, then encoded into numerical representations with Pre-trained Language Models (PLMs) \cite{ma2024knowlog, le2024prelog}, and finally consumed by diagnosis models for anomaly detection (AD) \cite{yang2021semi, le2021log}, root cause localization (RCL) \cite{wittkopp2024logrca, amar2019mining}, and fault identification (FI) \cite{duan2025famos, sui2023logkg}. 
While the former is effective for deep reasoning over a limited number of cases, it is less suitable as the high-throughput front end of production diagnosis pipelines, where systems must continuously encode, index, filter, and retrieve evidence from massive log streams under strict latency, deployment, cost, and privacy constraints \cite{ma2026loginstruct, xiao2025clslog, ma2025logreasoner}. In such settings, the latter remains indispensable.
In this paper, we focus on the representation stage in the second paradigm.

Despite its central role in diagnosis, existing PLM-based log representation methods still fail to meet the requirements of fault diagnosis \cite{llmelog, ma2024knowlog, le2024prelog}. 
LLMeLog \cite{llmelog} conducted an empirical study which proved that log representations encoded by PLMs suffer from issues such as content missing, semantic deflection, and tendency lacks, leading to a 19.97\% decrease in anomaly detection. 
KnowLog \cite{ma2024knowlog} pointed out that PLMs have significant deficiencies in understanding the abbreviations, context, and style of logs. 
PreLog \cite{le2024prelog} analyzed the inadequacy of PLMs in understanding log evolution. 
Overall, leveraging the full potential of PLMs for software system fault diagnosis still faces significant challenges.

In our view, the root cause of these issues is the inherent structural differences between logs and natural language. 
The characteristics of system faults do not exist in a singular semantic form but are manifested through the multi-level structure of logs. 
Therefore, relying solely on models pre-trained on natural language makes it difficult to comprehensively capture and understand this complex structural information, thus impeding effective fault diagnosis.

In fact, in real-world software system reliability maintenance practices, Site Reliability Engineers (SREs) typically analyze logs from three levels: execution flow, event, and component, to gain insights into system faults, as shown in Figure \ref{fig:teaser}. 
1) Execution Flow Level: SREs focus on scrutinizing execution flows that deviate from historical patterns. Such unexpected execution paths are considered strong indicators of nomalies \cite{du2017deeplog, meng2019loganomaly}. 
2) Event Level. In production practice, SREs abstract a single log entry into an independent system event. The specific semantics carried by each event serve as the key basis for them to determine the cause of a fault \cite{llmelog}. 
3) Component Level: Log records contain fine-grained information about system components (e.g., GPU, disk). By deeply analyzing this information, SREs can localize the fault to specific hardware or software units \cite{duan2025famos}.

We term the information that SREs obtain from the multi-level analysis of logs as fallibility representations. PLMs, operating in the form of masked language modeling, can only encode the natural language information of system logs while neglecting the fault-related fallibility representations, leading to poor performance of the encoded log representations in various fault diagnosis tasks. 
To investigate the capability of PLMs in encoding fallibility representations, this paper conducts an empirical study on existing PLMs using three log datasets. The results indicate that existing PLMs are indeed incapable of encoding these fallibility representations, which leads to unsatisfactory fault diagnosis performance.

Solving this problem requires comprehensively learning fallibility representations from system logs, which presents two challenges: 1) designing objectives to learn different types of fallibility, which is not directly observable; and 2) mitigating interference among their learning processes to learn them comprehensively.

To address these issues and unlock the potential of PLMs in fault diagnosis tasks, in this paper, we propose \textsc{Bifrost}, an innovative log representation learning method designed to enhance the capabilities of PLMs in fault diagnosis by learning fallibility representations from the continuously generated system logs of software systems. 
To tackle the first problem, \textsc{Bifrost} first meticulously designs three training tasks: Execution Flow Prediction (EFP), Abnormal Event Discrimination (AED), and System Component Perception (SCP). 
These tasks, based on the paradigm of self-supervised contrastive learning, are capable of learning different types of fallibility representations from system logs. 
To address the second problem, we first conducts an error analysis of the learning process for fallibility representations. 
We discovered that due to the inherent random sampling process in the fallibility learning process, an error is generated at each step of the optimization trajectory, which ultimately accumulates in the model's weight parameters, a phenomenon we term the fallibility jitter problem. 
To solve the fallibility jitter problem, \textsc{Bifrost} proposes the Co-Anchored Fallibility Representation Learning (CARL) strategy. The key idea of this strategy is to use system logs from real execution flows as representation space anchors for multi-objective representation learning, thereby reducing weight errors during the learning process.

To evaluate the effectiveness of \textsc{Bifrost}, we conducted exhaustive evaluations on three of the most widely used public datasets \cite{zhu2023loghub, oliner2007supercomputers} for log-based fault diagnosis—BGL, Hadoop, and Thunderbird—as well as on dataset collected from an industrial ML-as-a-Service (MLaaS) system, Platform-X, covering the three fault diagnosis tasks of AD, RCL, and FI. 
The experimental results demonstrate that the log representations generated by \textsc{Bifrost} achieve advanced performance in fault diagnosis tasks. 
Compared to PLMs with a similar number of parameters, \textsc{Bifrost} achieved an average advantage of 20.81\% across the three tasks. 
Compared to several larger baselines, \textsc{Bifrost} also shows competitive performance.

In summary, our contributions are as follows: 
\begin{itemize}
    \item We analyzed the log analysis experience of SREs and identified three types of information closely related to failure, which we term fallibility representations.
    \item We conducted an empirical study, the results of which show that existing PLM-based log representation methods perform poorly in encoding fallibility representations, leading to suboptimal failure diagnosis.
    \item We propose \textsc{Bifrost}, a representation learning method that enhances the failure diagnosis capabilities of PLMs by learning fallibility representations.
    \item Quantitative, qualitative, and evaluations on four datasets validate \textsc{Bifrost} and fallibility representations for diagnosis.
\end{itemize}

\section{Task Formulation}

Our objective is log-based fault diagnosis for software systems. Consider a large-scale software system. Within an observation window of length $w$, we define the log sequence as $\mathbf{x}=[l_1,l_2,\ldots,l_w]$, where $l_i$ is a log event observed at a time step. Fault diagnosis aims to establish a framework that processes the input log sequence and outputs 1) the system’s operational status and, when an anomaly is detected, 2) the root-cause logs and 3) the fault type. This task consists of three sub-tasks:
\begin{itemize}[leftmargin=*]
    \item \textbf{Anomaly Detection (AD)} \cite{huang2025codead, midlog, llmelog}: Given an observed log sequence $\mathbf{x}$, a detector $\mathcal{D}$ is used to predict the presence of an anomaly. Its output, $y$, is a binary variable indicating whether the system is normal (0) or anomalous (1), i.e., $y=\mathcal{D}(\mathbf{x})\in\{0,1\}$.
    \item \textbf{Root Cause Localization (RCL)} \cite{zhang2025adaptive, wang2020root, rosenberg2020spectrum}: If the system state is determined to be anomalous, a localizer $\mathcal{R}$ is activated to evaluate the probability of each log $l_i$ in the input sequence $\mathbf{x}$ being the root cause. That is, $\mathcal{R}(\mathbf{x})=[p_1,p_2,\ldots,p_w]$, where $p_i\in[0,1]$ represents the probability that log $l_i$ is the root cause.
    \item \textbf{Fault Identification (FI)} \cite{duan2025famos, sui2023logkg}: After confirming a system anomaly, a multi-class classifier $\mathcal{M}$ is employed to identify the specific type of the fault. It is assumed that the potential fault types belong to a predefined, finite set $\mathcal{U} = \{u_1,u_2,\ldots,u_{|\mathcal{U}|}\}$. The objective of the $\mathcal{M}$ is to predict the fault type corresponding to the current log sequence, i.e., $\mathcal{M}(\mathbf{x})\in \mathcal{U}$.
\end{itemize}

\section{Empirical Study}
\label{sec:empirical}
In section, we conduct an empirical study on existing PLM-based log representation methods on three systems \cite{zhu2023loghub, oliner2007supercomputers}: Thunderbird, BGL, and an industrial MLaaS system Platform-X. 
We manually examined the capability of the four most widely used PLMs \cite{pennington2014glove, lanalbert, devlin2019bert, lewis2020bart} to encode fallibility representations. 
\textbf{Section \ref{sec:appendix_exp}} provides a detailed introduction to the baseline models and the systems.

\subsection{How effective are existing PLMs in identifying unexpected execution flows?}
In this section, we evaluate the capability of existing PLMs to identify unexpected execution flows in the form of predicting the next event.
We first collect a normal execution flow of the system, denoted as $L=[l_1, l_2, \ldots, l_T]$. Then, we use a time window $w$ to partition $L$ into execution flows of length $w$, represented as $\mathbf{x}_t=[l_{t-w}, \ldots, l_{t-1}]$.
We use PLMs, denoted as $f_{\theta}(\cdot)$, for encoding to obtain a sequence of log representations $f_{\theta} (\mathbf{x}_t)$. This sequence is subsequently fed into a sequential neural network $g_\phi(\cdot)$ to predict the next log in the execution flow: $\hat{e}_{t} = g_\phi(f_{\theta} (\mathbf{x}_t))$.
Our training objective is to maximize the cosine similarity $s(\hat e_{t}, e_{t})$ between $\hat{e}_{t}$ and the embedding of the next true log in the execution flow, $e_{t} = f_{\theta} (l_{t})$. 
Here, the parameters $\theta$ of the PLM $f_{\theta}$ are frozen to evaluate the representation quality of the original PLMs.
After training is complete, we use unexpected execution flows for evaluation. 
For an unexpected event representation $\tilde e_{t}$, a smaller value of $s(\hat e_{t}, \tilde{e}_{t})$ indicates a stronger capability of the PLM to identify unexpected execution flows.

We use the absolute cosine similarity, $Avg\_Sim$, and the relative rank similarity, $Avg\_Rank$, to quantify the similarity:
\begin{itemize}
    \item $Avg\_Sim = \mathbb{E} \left[ s(\tilde e_i, \hat{e}_i)\right]$
    \item $Avg\_Rank = \mathbb{E} \left[ 1 + \sum_{e_j \in \mathbf{E} (L)} \mathbb{I}\left(s(\hat{e}_i, \tilde{e}_i) \le s(\hat{e}_i, e_j)\right) \right]$
\end{itemize}
where $\mathbb{E}$ denotes the expected value and $\mathbb{I}$ is the indicator function. A lower $Avg\_Sim$ suggests a stronger ability of the PLM to identify unexpected execution flows, whereas for $Avg\_Rank$, the opposite is true.
The experimental results on the Thunderbird system logs are shown in Table \ref{tab:empirical_ad}. It is readily apparent that existing methods consider unexpected execution flows to have a high similarity to historical execution flows, which indicates that the capability of existing PLMs to identify unexpected execution flows is not ideal.

\begin{table}[htbp]
  \centering
  \captionsetup{skip=2pt}
  \caption{Representational similarity between expected and unexpected events in PLMs.}
    \scalebox{0.9}{
    \begin{tabular}{ccc}
    \toprule
    \textbf{Embedding Model} & \multicolumn{2}{c}{\textbf{Thunderbird}} \\
    \midrule
    \textbf{Metrics} & \textbf{Avg\_Sim $\downarrow$} & \textbf{Avg\_Rank $\uparrow$} \\
    \midrule
    Glove-300d \cite{pennington2014glove} & 78.08 & 13.49 \\
    BERT-Base \cite{devlin2019bert} & 89.62 & 21.69 \\
    ALBERT-Base \cite{lanalbert} & 92.22 & 21.05 \\
    BART-Base \cite{lewis2020bart} & 81.03 & 17.16 \\
    \midrule
    Average & 88.10 & 20.94 \\    
    \bottomrule
    \end{tabular}%
    }
  \label{tab:empirical_ad}
  \vspace{-0.4cm}
\end{table}

\subsection{Can existing PLMs discriminate anomalous system events?}
\label{sec:empirical_rca}
\begin{figure}[htbp]
    \centering
    \begin{subfigure}[b]{0.45\linewidth}
        \includegraphics[width=\linewidth]{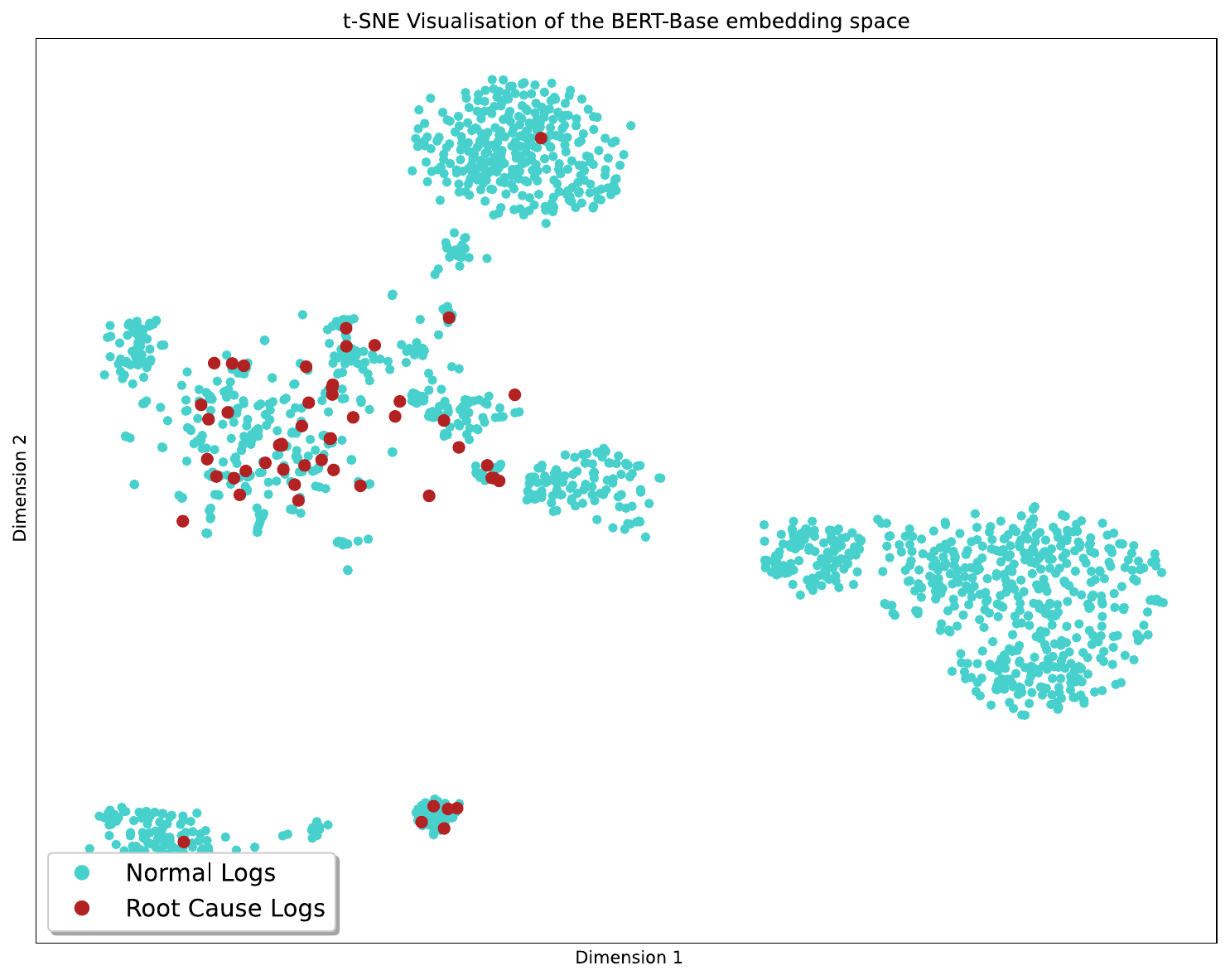}
        \caption{BERT-Base}
        \label{bert_log}
    \end{subfigure}
    \hfill
    \begin{subfigure}[b]{0.45\linewidth}
        \includegraphics[width=\linewidth]{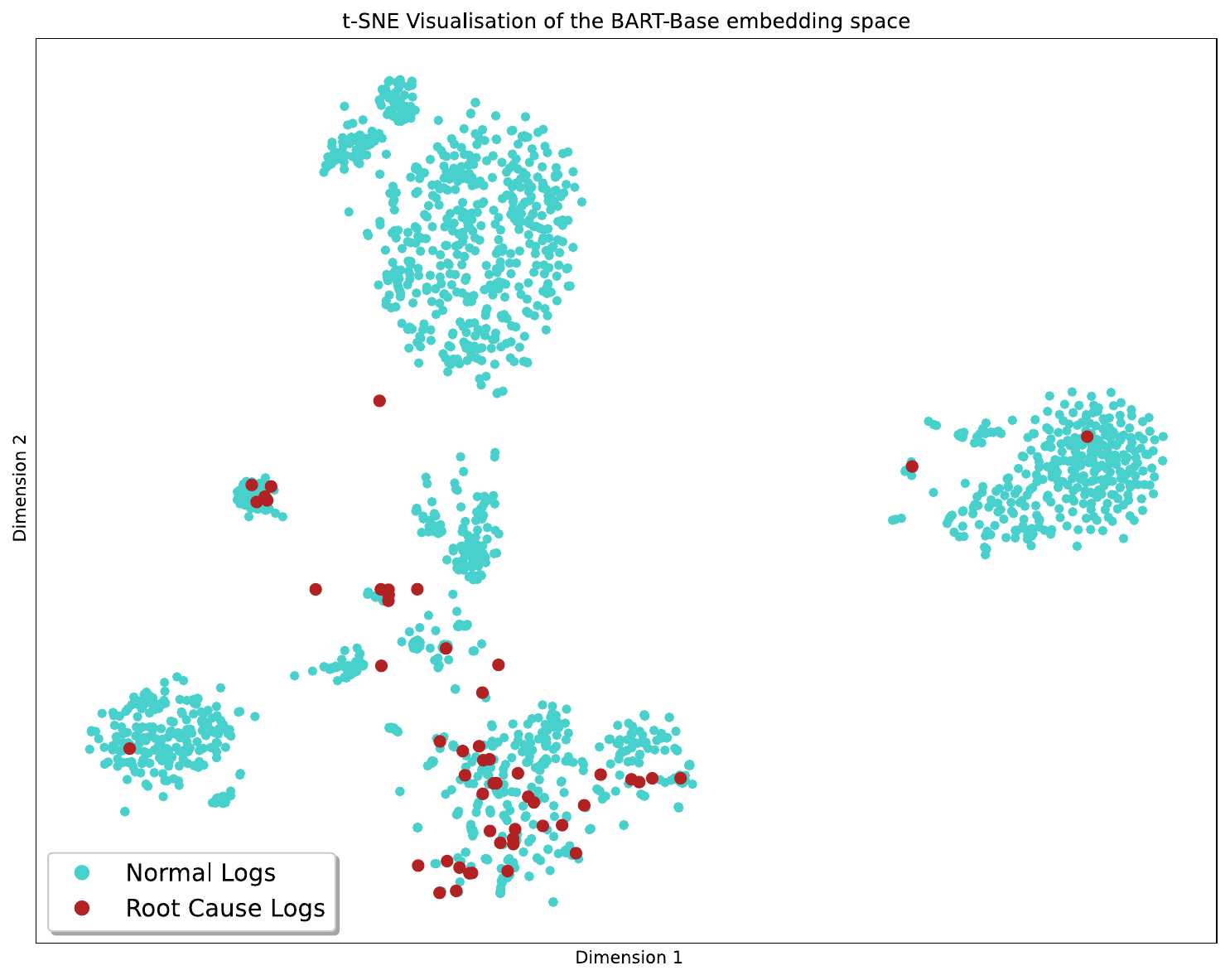}
        \caption{BART-Base}
        \label{bart_log}
    \end{subfigure}
    \setlength{\abovecaptionskip}{-1pt}
    \caption{\small{System-event Representations on BGL generated by PLMs.}}
    \Description{System-event Representations on BGL generated by PLMs.}
    \label{fig:log_repr}
    \vspace{-0.4cm}
\end{figure}

This section investigates the discriminative ability of existing PLMs for anomalous system events. Limited by space constraints, we select the most representative BERT-Base \cite{devlin2019bert} and BART-Base \cite{lewis2020bart} as baseline models, and choose the BGL system for our study. We use PLMs to encode the logs from the BGL system and then visualize these representations using t-SNE in Figure \ref{bert_log} and Figure \ref{bart_log}. 

It is easy to observe that in the log representations obtained by existing PLMs, normal system events and anomalous system events are heavily stacked together, and the distribution of anomalous system events is relatively dispersed. This evidence indicates that existing PLMs do not have the ability to distinguish anomalous semantics within logs, resulting in non-ideal log representations.

\begin{figure}[htbp]
    \centering
    \begin{subfigure}[b]{0.45\linewidth}
        \includegraphics[width=\linewidth]{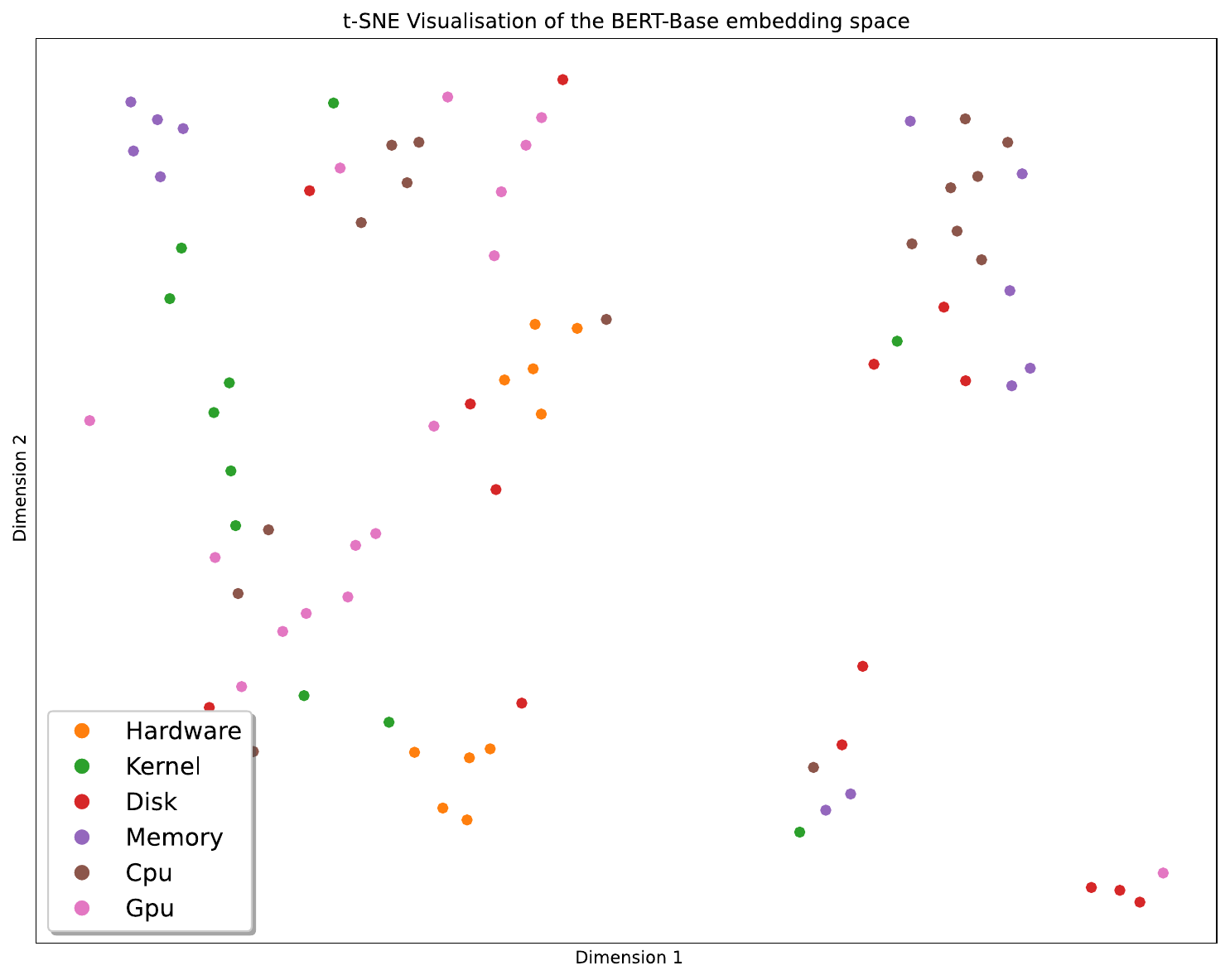}
        \caption{BERT-Base}
        \label{bert_component}
    \end{subfigure}
    \hfill
    \begin{subfigure}[b]{0.45\linewidth}
        \includegraphics[width=\linewidth]{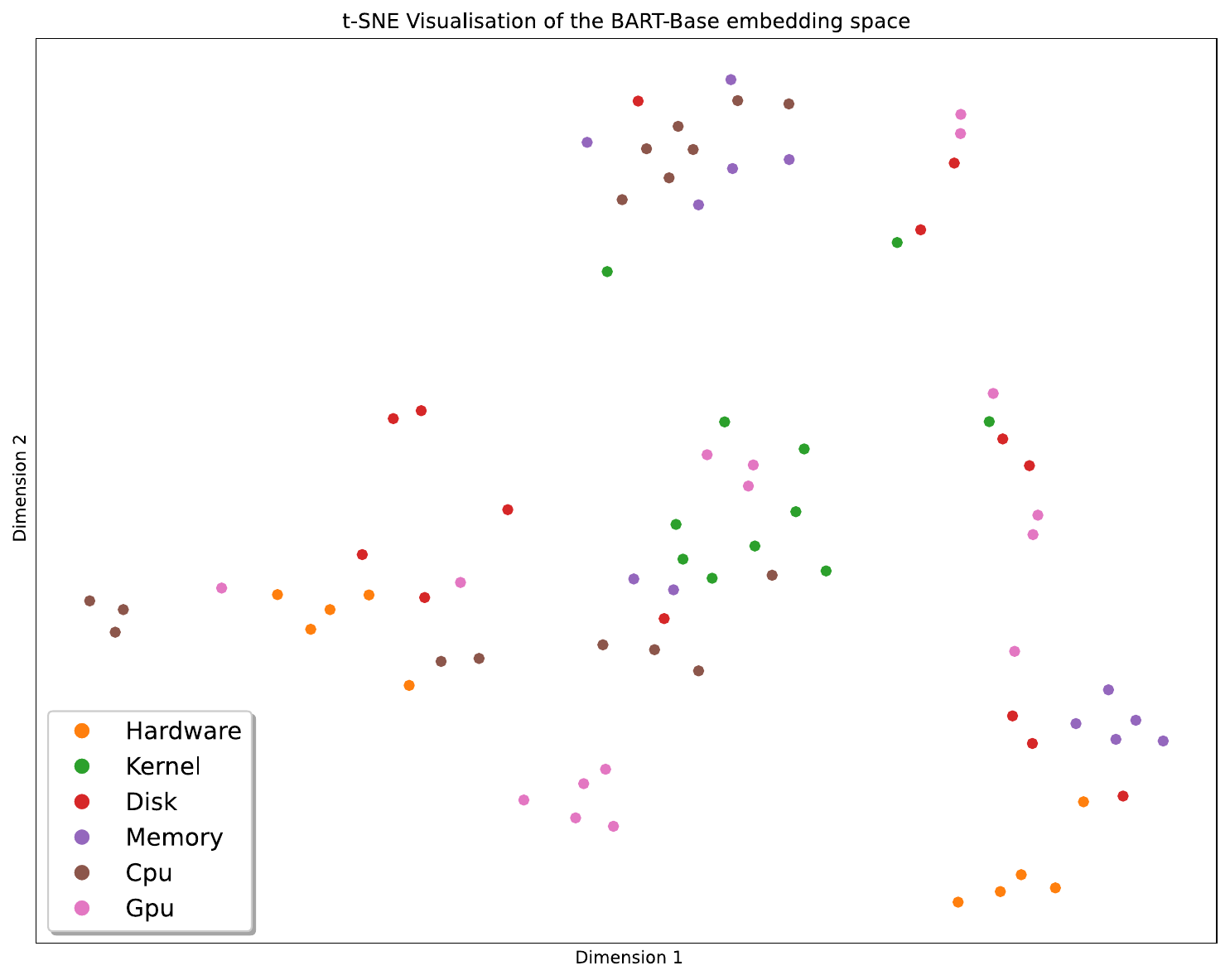}
        \caption{BART-Base}
        \label{bart_component}
    \end{subfigure}
    \setlength{\abovecaptionskip}{-1pt}
    \caption{\small{System-component Representations generated by PLMs.}}
    \Description{System-component Representations generated by PLMs.}
    \label{fig:component_repr}
    \vspace{-0.4cm}
\end{figure}

\subsection{Can existing PLMs perceive different system components?}
This section investigates the capability of existing PLMs to perceive different system components. Due to space constraints, we select the most representative BERT-Base and BART-Base as baseline models and choose Platform-X as the system for our study. Similar to Section \ref{sec:empirical_rca}, we use PLMs to encode the logs from the Platform-X system and then visualize these representations using t-SNE, with the results shown in Figure \ref{bert_component} and \ref{bart_component}.
It is easy to observe that in the log representations obtained by existing PLMs, the representations of different system components are stacked together. This indicates that it is difficult for existing PLMs to perceive different system components from the raw logs.

\begin{center}
    \begin{tcolorbox}[colback=gray!10,
        colframe=black,
        width=\linewidth,
        arc=1mm, auto outer arc,
        boxrule=0.5pt,
        top=2pt,
        bottom=2pt,
        left=2pt,
        right=2pt
        ]
        \small \textbf{Summary.} PLM-based log representation methods are only pre-trained on natural language and overlook the multi-level structure of logs, leading to: 1) an inability to identify unexpected execution flows, 2) insensitivity to anomalous system events, and 3) a failure to separate different system components. 
        Thus resulting in suboptimal fault diagnosis performance.
    \end{tcolorbox}
\end{center}

\section{Methodology}\label{sec:method}

\subsection{Overview}

\begin{figure*}[htbp]
\centering
\captionsetup{skip=2pt}
\includegraphics[width=0.95\linewidth]{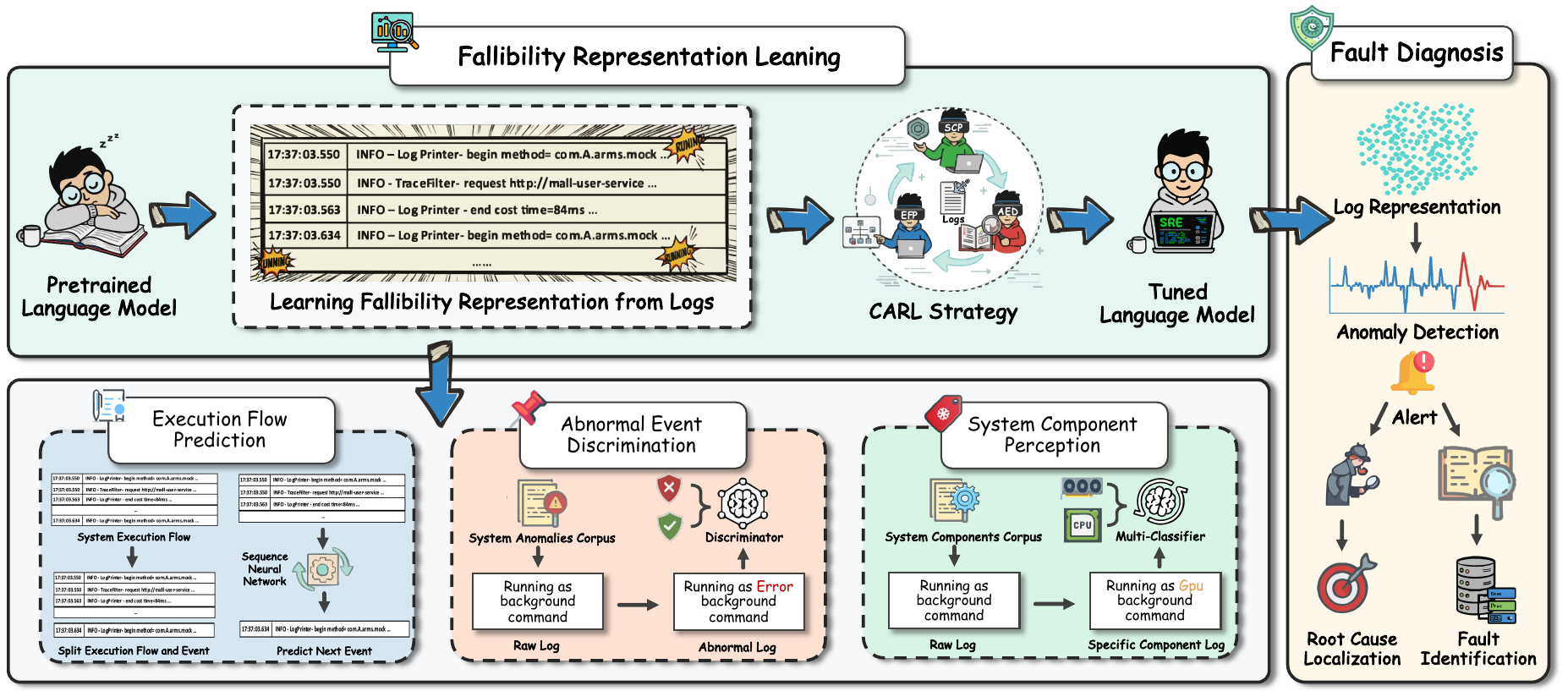}
\caption{\small \textsc{Bifrost} is initialized from the PLM BART-Base. It is then trained via a carefully designed representation learning process, which learns log fallibility representations from logs collected during normal system operation. After training, \textsc{Bifrost} encodes logs for fault diagnosis.}
\Description{\textsc{Bifrost} is initialized from the PLM BART-Base. It is then trained via a carefully designed representation learning process, which learns log fallibility representations from logs collected during normal system operation. After training, \textsc{Bifrost} encodes logs for fault diagnosis.}
\label{fig:pipeline}
\vspace{-0.4cm}
\end{figure*}

This paper introduces \textsc{Bifrost}, an innovative log representation learning method designed to enhance the capabilities of PLMs in fault diagnosis. While log-based fault diagnosis is crucial for maintaining the reliability of software systems, existing PLM-based methods primarily encode semantic information from logs. They often fail to perceive system failure patterns embedded within the multi-level structure of logs, leading to suboptimal diagnostic performance. To bridge this gap, Bifrost draws inspiration from the analytical strategies of SREs. \textsc{Bifrost} formulate three distinct learning objectives, each targeting a specific type of fallibility representation corresponding to a different level of the log structure. 
However, influenced by in-batch stochastic sampling, the optimization process of fallibility learning exhibits significant instability. We term this \textbf{Fallibility Jitter} and have analyzed the weight error upper bound.
To address the fallibility jitter problem, \textsc{Bifrost} proposes the \textbf{Co-Anchored Fallibility Representation Learning} strategy. 
By robustly learning these representations, \textsc{Bifrost} produces log embeddings that are highly effective for fault diagnosis tasks. 
The overall pipeline of \textsc{Bifrost} is illustrated in Figure \ref{fig:pipeline}.

\subsection{Fallibility Representation for Fault Diagnosis}

\textsc{Bifrost} learns \textbf{fallibility representations} from logs to produce embeddings tailored for fault diagnosis. 
Since such representations are latent and cannot be directly supervised, we operationalize fallibility learning with three self-supervised contrastive objectives: Execution Flow Prediction (EFP), Abnormal Event Discrimination (AED), and System Component Perception (SCP).

\subsubsection{\textbf{Self-Supervised Learning of Fallibility Representations}}

Since fallibility is a latent property spanning multiple log-structure levels, direct supervision is difficult to obtain.
We view fallibility as \emph{relational} in the embedding space: instances preserving level-specific consistency should be close, whereas those violating it should be separated.
For example, logs following a normal execution flow should cluster, while logs linked to behaviors, anomalous event semantics, or different system components should be farther apart.
This motivates a self-supervised contrastive objective \cite{khosla2020supervised, chuang2020debiased} that provides proxy supervision for learning fallibility representations.

Let a chronological log stream from normal operation be $\mathbf{L}=[l_1,\ldots,l_T]$.
An embedding model $f_{\theta}(\cdot)$ maps a log message or a log sequence to $\mathbb{R}^{d}$.
We use cosine similarity $s(\cdot,\cdot)$ and temperature hyperparameter $\tau$.
We use the term \emph{instance} to denote a training input (a log message or a context window).
Given an anchor instance $v$, we sample a positive instance $v^{+}\in\mathcal{P}(v)$ that preserves level-specific consistency, and sample $K$ negative instances $\{\tilde{v}_k\}_{k=1}^{K}\subset\mathcal{N}(v)$ that violate this consistence.
We then define the InfoNCE loss \cite{oord2018representation}:

\begin{equation}
\scalebox{0.9}{
    $
\begin{aligned}
&\mathcal{L}_{\text{CL}}\big(v, v^{+}, \mathcal{N}(v)\big) \\
&= - \log \frac{\exp(s(f_{\theta}(v), f_{\theta}(v^{+})) / \tau)}
{\exp(s(f_{\theta}(v), f_{\theta}(v^{+})) / \tau) + \sum_{\tilde{v} \in \mathcal{N}(v)} \exp(s(f_{\theta}(v), f_{\theta}(\tilde{v})) / \tau)} .
\end{aligned}
    $
}
\end{equation}

We minimize the mini-batch average of $\mathcal{L}_{\text{CL}}$ to optimize $\theta$.
Here, $\mathcal{P}(\cdot)$ and $\mathcal{N}(\cdot)$ specify instance construction rules that preserve or violate the target level-specific consistency.
Using this shared template, we instantiate three contrastive objectives aligned with the multi-level log structure: EFP, AED, and SCP.
Each objective defines the anchor instance $v$ and the corresponding $\mathcal{P}(\cdot)$ and $\mathcal{N}(\cdot)$ for a specific level of fallibility.

\subsubsection{\textbf{Execution Flow Prediction (EFP)}}
In real-world practice, SREs pay close attention to execution flows that deviate from established historical patterns. Such unexpected execution paths are considered strong indicators of system anomalies. Consequently, the ideal log representation must be capable of identifying these deviations. To this end, we propose the EFP task, which learns the patterns of normal system execution flows in a predictive manner.

To assess whether an upcoming event is compatible with the normal execution flow, EFP uses the preceding execution context as the anchor.
At time $t$, we form a context-window instance $\mathbf{x}_t=[l_{t-w},\ldots,l_{t-1}]$ and obtain its embedding via $f_{\theta}(\cdot)$.
An EFP-specific prediction head $g_{\phi}(\cdot)$, implemented as a sequence model, projects the context embedding into the log embedding space:
\begin{equation}
\hat e_t = g_{\phi}(f_{\theta}(\mathbf{x}_t)).
\end{equation}
The predictive anchor $\hat e_t$ summarizes the historical flow and is used to assess whether the subsequent event follows the normal execution flow.

To avoid template memorization, we sample a random in-window message $l_i\in\mathbf{x}_t$ and construct positives as its semantics-preserving perturbations, reflecting the flow state rather than the exact next message $l_t$.
Let $\mathcal{P}_{\text{EFP}}(l_i)$ be the set of variants produced by lightweight perturbations (random word deletion, synonym replacement, and adjacent word swapping), and we sample $l_i^{+}\in\mathcal{P}_{\text{EFP}}(l_i)$ as the positive instance.
This mitigates template overfitting, and improves generalization to unseen normal execution flows \cite{jing2023contrastive, liu2021self}.

To expose execution-flow violations without explicit labels, we contrast the predictive anchor against negative continuations that break execution-flow consistency.
Given a context $\mathbf{x}_t$ from normal operation, the observed next event $l_t$ serves as a flow-consistent continuation.
We construct negatives by injecting failure semantics from a fault lexicon $\mathcal{A}$ into $l_t$, yielding counterfactual continuations that introduce failure cues and are unlikely under the normal execution flow.
Specifically, we sample $K$ negatives $\{\tilde{l}_{t,k}\}_{k=1}^{K} \subset \mathcal{N}_{\text{EFP}}(l_t)$ by inserting fault phrases $a_k \in \mathcal{A}$ into $l_t$, where $\oplus$ inserts $a_k$ at a random position and $\mathcal{A}$ is described in Sec.~\ref{sec:fault-lexicon}.
\begin{equation}
\tilde{l}_{t,k} = l_t \oplus a_k,\quad a_k \in \mathcal{A},
\end{equation}
EFP then optimizes:
\begin{equation}
\scalebox{0.9}{
    $
\begin{aligned}
&\min_{\theta}\ \mathcal{L}_{\text{EFP}}\big(\hat e_t, l_i^{+}, \{\tilde{l}_{t,k}\}_{k=1}^{K}\big)
= \\
&-\frac{1}{T-w}\sum_{t=w+1}^{T}
\log \frac{\exp(s(\hat e_t, f_{\theta}(l_i^{+}))/\tau)}
{\exp(s(\hat e_t, f_{\theta}(l_i^{+}))/\tau) + \sum_{k=1}^{K}\exp(s(\hat e_t, f_{\theta}(\tilde{l}_{t,k}))/\tau)}.
\end{aligned}
    $
}
\end{equation}
This objective aligns $\hat e_t$ with flow-consistent evidence and separates it from flow-violating counterfactuals, thereby encoding execution-flow level-specific consistency in the representations.

\subsubsection{\textbf{Abnormal Event Discrimination (AED)}}

At the event level, SREs abstract individual log entries into discrete system events and reason about their semantics to identify the nature and cause of faults.
Thus, an effective log representation should also separate normal and abnormal semantics at the event level.
To this end, \textsc{Bifrost} includes the AED objective, which directly contrasts normal event semantics against anomaly-like variants.

In AED, each log message from normal operation $l_t \in \mathbf{L}$ is treated as an event-level anchor instance.
For a given anchor $l_t$, we construct a positive instance $l_t^{+} \in \mathcal{P}_{\text{AED}}(l_t)$ by applying the same semantics-preserving perturbations as in EFP.
These perturbations preserve the event intent while encouraging robustness to superficial lexical changes, so that variants of the same normal event cluster in the embedding space. To model abnormal semantics, we construct negative instances by injecting failure-related phrases from the fault lexicon $\mathcal{A}$ into the anchor event.
Specifically, we sample $K$ negatives $\{\tilde{l}_{t,k}\}_{k=1}^{K} \subset \mathcal{N}_{\text{AED}}(l_t)$ of the form
\begin{equation}
\tilde{l}_{t,k} = l_t \oplus a_k,\quad a_k \in \mathcal{A},
\end{equation}
where $\oplus$ inserts $a_k$ at a random position and $\mathcal{A}$ is described in Sec.~\ref{sec:fault-lexicon}.
This injection yields counterfactual variants that introduce failure cues and are unlikely under normal operation.
Given the anchor $l_t$, the positive $l_t^{+}$, and the negatives $\{\tilde{l}_{t,k}\}_{k=1}^{K}$, AED instantiates the contrastive loss as:
\begin{equation}
\scalebox{0.8}{
    $
\begin{aligned}
&\min_{\theta}\ \mathcal{L}_{\text{AED}}\big(l_t, l_t^{+}, \{\tilde{l}_{t,k}\}_{k=1}^{K}\big)
= \\
&-\frac{1}{{T-w}}\sum_{t=w+1}^{T}
\log \frac{\exp(s(f_{\theta}(l_t), f_{\theta}(l_t^{+}))/\tau)}
{\exp(s(f_{\theta}(l_t), f_{\theta}(l_t^{+}))/\tau) + \sum_{k=1}^{K}\exp(s(f_{\theta}(l_t), f_{\theta}(\tilde{l}_{t,k}))/\tau)}.
\end{aligned}
    $
}
\end{equation}
This objective pulls embeddings of semantics-preserving variants of a normal event together and pushes them away from failure-injected variants, thereby capturing event-level fallibility.

\subsubsection{\textbf{System Component Perception (SCP)}}
\label{sec:scp}

Log entries often contain explicit references to system components (e.g., GPU, disk, memory).
By analyzing such component cues, SREs can localize faults to specific software or hardware units and narrow the diagnostic scope.
Therefore, effective log representations should also be sensitive to component identity.
The SCP objective is designed to encode this component-level fallibility in the embedding space.

We assume a predefined component vocabulary $\mathcal{C} = \{c_1, \ldots, c_{|\mathcal{C}|}\}$ that covers common component types in distributed systems (e.g., \texttt{cpu}, \texttt{memory}), the full list is given in Appendix~\ref{app:system-components}.
This reflects the practical observation that many failures manifest at the component granularity, and these keywords provide a simple and portable inductive bias.
For each log message $l_t$ from normal operation, we randomly sample a component keyword $c_j \in \mathcal{C}$ and construct an anchor instance by inserting $c_j$ into $l_t$:
\begin{equation}
l_{t,j} = l_t \oplus c_j,
\end{equation}
where $\oplus$ denotes string insertion at a random position.
Thus, $l_{t,j}$ can be viewed as a log instance explicitly associated with component $c_j$.

To encourage component-consistent clustering and robustness to surface variations, we construct positives as semantics-preserving variants of the anchor.
Specifically, we define $\mathcal{P}_{\text{SCP}}(l_{t,j})$ as augmentations obtained by applying the lightweight perturbations from EFP and AED to $l_{t,j}$, and sample a positive instance $l_{t,j}^{+} \in \mathcal{P}_{\text{SCP}}(l_{t,j})$.
These augmentations keep the component keyword $c_j$ while making representations invariant to minor lexical changes.

To separate different components, we construct negative instances by injecting alternative component keywords into the same base log.
For each anchor $l_{t,j}$, we randomly sample $Z$ negative instances $\{\tilde{l}_{t,j,z}\}_{z=1}^{Z} \subset \mathcal{N}_{\text{SCP}}(l_t, c_j)$ of the form
\begin{equation}
\tilde{l}_{t,j,z} = l_t \oplus c_i,\quad c_i \in \mathcal{C} \setminus \{c_j\},
\end{equation}
again using $\oplus$ to denote string insertion.
These negatives are logs sharing the same base content but associated with different components and should be distinguished. Given the anchor $l_{t,j}$, the positive $l_{t,j}^{+}$, and the negatives $\{\tilde{l}_{t,j,z}\}_{z=1}^{Z}$, SCP optimizes:
\begin{equation}
\scalebox{0.8}{
    $
\begin{aligned}
&\min_{\theta}\ \mathcal{L}_{\text{SCP}}\big(l_{t,j}, l_{t,j}^{+}, \{\tilde{l}_{t,j,z}\}_{z=1}^{Z}\big)
= \\
&-\frac{1}{{T-w}}\sum_{t=w+1}^{T}
\log \frac{\exp(s(f_{\theta}(l_{t,j}), f_{\theta}(l_{t,j}^{+}))/\tau)}
{\exp(s(f_{\theta}(l_{t,j}), f_{\theta}(l_{t,j}^{+}))/\tau) + \sum_{k=1}^{K}\exp(s(f_{\theta}(l_{t,j}), f_{\theta}(\tilde{l}_{t,j,z}))/\tau)},
\end{aligned}
    $
}
\end{equation}
This objective pulls together instances associated with the same component keyword and pushes them away from instances associated with different components, thereby capturing component-level fallibility in the learned representations.

\subsubsection{\textbf{Fault Lexicon Construction}}
\label{sec:fault-lexicon}
The negative generators in EFP and AED inject failure semantics by inserting phrases from a fault lexicon $\mathcal{A}$.
A manually crafted lexicon is labor-intensive to maintain and may not reflect how failures are described in the target logs.
We therefore construct $\mathcal{A}$ from the training corpus via two stage: \emph{candidate recall} followed by \emph{reranking}.

\textbf{Stage 1: Candidate Recall.}
The goal of this stage is to retrieve a broad set of corpus-specific candidate expressions associated with failure semantics in system logs.
Following standard log-level conventions in SRE, we use tokens such as \texttt{warning}, \texttt{error}, and \texttt{fatal} as seed anchors.
We tokenize the training logs and WordNet \cite{miller1995wordnet} and merge them into a unified retrieval vocabulary.
For each anchor, we embed all its instances using Qwen3-Embedding-0.6B \cite{yang2025qwen3} and use the mean vector as the query representation.
We then embed each entry in the retrieval vocabulary, rank entries by cosine similarity to the query, and keep the top 300 as the candidate pool $\Omega$.

\textbf{Stage 2: Reranking.}
High-frequency phrases are not necessarily fault-related, many correspond to generic descriptions.
To focus on failure-related semantics, we employ a reranker $r_{\psi}(q, \kappa)\in[0,1]$ that scores the semantic relatedness between an anchor query $q$ and a candidate phrase $\kappa\in \Omega$.
We set $q$ to the concatenation of the three log-level anchors, \texttt{warning, error, fatal}, score each candidate in $\Omega$, and select the top-$m$ phrases as the final fault lexicon:
\begin{equation}
\mathcal{A} = \operatorname{Top}_{m}\big(\{\kappa \in \Omega \},\, r_{\psi}(q,\kappa)\big).
\end{equation}
We instantiate $r_{\psi}(\cdot,\cdot)$ with Qwen3-Reranker-0.6B \cite{yang2025qwen3} and set $m{=}30$ by default.
This pipeline yields a compact, corpus-grounded lexicon of failure-related phrases, supporting realistic and diverse negative instance construction for fallibility learning.

\subsection{Learning from Fallibility Representation}

The goal of \textsc{Bifrost} is to provide an end-to-end framework for fallibility representation learning. 
We first analyze a vanilla training strategy that jointly optimizes the three objectives.
We find that inherent stochastic sampling induces significant optimization instability, a phenomenon we term \textbf{fallibility jitter}, which prevents the model from converging to an ideal representation space.
To mitigate this issue, we propose the \textbf{CARL} strategy, which enables more robust and efficient representation learning by anchoring optimization on real system events.

\subsubsection{\textbf{Fallibility Jitter in Representation Learning}}

To learn fallibility representations in an end-to-end manner, the vanilla strategy jointly optimizes the three objectives.
At optimization step $s$, we sample a time index $t$ and denote the log at time $t$ by $l_t$, together with its context window $\mathbf{x}_t = [l_{t-w}, \ldots, l_{t-1}]$, a log message $l_i \in \mathbf{x}_t$ for EFP, and a component keyword $c_j \in \mathcal{C}$, and construct the SCP anchor as $l_{t,j} = l_t \oplus c_j$.
The vanilla loss is:
\begin{equation} \label{eq:vanilla_loss}
\scalebox{1}{
    $
\begin{aligned}
\mathcal{L}_{\text{Vanilla}}
&= \lambda_1 \mathcal{L}_{\text{EFP}}(\hat e_t, l_i^{+}, \{\tilde l_{t,k}\}_{k=1}^{K})
+ \lambda_2 \mathcal{L}_{\text{AED}}(l_t, l_t^{+}, \{\tilde l_{t,k}\}_{k=1}^{K})
\\&+ \lambda_3 \mathcal{L}_{\text{SCP}}(l_{t,j}, l_{t,j}^{+}, \{\tilde l_{t,j,z}\}_{z=1}^{Z}),
\end{aligned}
    $
}
\end{equation}

We provide a theoretical analysis of the optimization process for $\mathcal{L}_{\text{Vanilla}}$ in \textbf{Appendix~\ref{sec:fj_define}}.
Our analysis shows that inherent stochastic sampling causes the parameter trajectory $\mathbf{W}'_s$ driven by stochastic gradients to deviate from the ideal trajectory $\mathbf{W}_s$ driven by the true gradient, leading to suboptimal representations.
We refer to this phenomenon as \textbf{fallibility jitter}.
In \textbf{Appendix~\ref{sec:bound_all}}, we further derive an upper bound on the resulting parameter error:
\begin{equation} \label{eq:vanilla_error}
\mathbb{E}[\lVert \mathbf{W}_S - \mathbf{W}'_S \rVert] \le \eta \sum_{k=0}^{S-1} (1+\eta L_{\text{Lip}})^{S-1-s} \mathbb{E}_{\mathbf{W}'_s}[\sigma_{\text{Vanilla}}(\mathbf{W}'_s)],
\end{equation}
where $\sigma_{\text{Vanilla}}$ is the single-step gradient standard deviation, $\eta$ is the learning rate, $L_{\text{Lip}}$ is the Lipschitz constant of the gradient function, and $S$ is the total number of optimization steps. 
Equation~(\ref{eq:vanilla_error}) shows that the gradient variance $\sigma_{\text{Vanilla}}^2$ at each step is amplified and accumulated into the final parameter deviation, causing the parameters to oscillate along the convergence path and degrading the quality of fallibility representation learning.
This deviation is fundamentally induced by the stochastic sampling inherent in the vanilla optimization process.

\subsubsection{\textbf{Co-Anchored Fallibility Representation Learning}}

To mitigate fallibility jitter, we propose the CARL strategy.
The key idea is to use the real system event $l_t$ as a common optimization anchor, providing a deterministic center for each stochastic batch and stabilizing the training process.
At each optimization step $s$, we sample a time index $t$ and denote the log at time $t$ by $l_t$, together with its context window $\mathbf{x}_t = [l_{t-w}, \ldots, l_{t-1}]$, a context log $l_i \in \mathbf{x}_t$ for EFP, and a component keyword $c_j \in \mathcal{C}$, and construct the SCP anchor as $l_{t,j} = l_t \oplus c_j$.
CARL then redesigns the joint objective as:
\begin{equation}\label{eq:carl_loss}
\scalebox{0.9}{
    $
\begin{aligned}
\mathcal{L}_{\text{CARL}} &=
\lambda_1 \mathcal{L}_{\text{EFP}}(\hat e_t, l_t^{+}, \{\tilde n_{t,j,k}\}_{k=1}^{K+Z}\big)
+ \lambda_2 \mathcal{L}_{\text{CL}}(l_t^{+}, l_i^{+}, \{\tilde n_{t,j,k}\}_{k=1}^{K+Z})
\\&\quad+ \lambda_3 \mathcal{L}_{\text{AED}}(l_t, l_t^{+}, \{\tilde n_{t,j,k}\}_{k=1}^{K+Z})
+ \lambda_4 \mathcal{L}_{\text{SCP}}(l_{t,j}, l_{t,j}^{+}, \{\tilde l_{t,j,z}\}_{z=1}^{Z}),
\end{aligned}
    $
}
\end{equation}
\begin{equation}
\scalebox{1.0}{
    $
\begin{aligned}
\{\tilde n_{t,j,k}\}_{k=1}^{K+Z} \subset \{\,l_t \oplus a \mid a \in \mathcal{A}\,\} \cup \{\,l_t \oplus c \mid c \in \mathcal{C} \setminus \{c_j\}\,\}, 
\end{aligned}
    $
}
\end{equation}

\begin{equation}
\scalebox{1.0}{
    $
\begin{aligned}
\{\tilde l_{t,j,z}\}_{z=1}^{Z} \subset \{\,l_t \oplus c \mid c \in \mathcal{C} \setminus \{c_j\}\,\}.
\end{aligned}
    $
}
\end{equation}

The key design choices are:
(1) a revised EFP term that uses the central event $l_t^{+}$ as the flow-consistent continuation (the $\lambda_1$ term);
(2) an auxiliary contrastive constraint $\mathcal{L}_{\text{CL}}$ between $l_t^{+}$ and the sampled context log $l_i^{+}$ to explicitly handle the stochasticity from context sampling (the $\lambda_2$ term); and
(3) a shared negative set $\{\tilde n_{t,j,k}\}_{k=1}^{K+Z}$ across EFP, $\mathcal{L}_{\text{CL}}$, and AED, which couples their gradients through common negative evidence.
We analyze the optimization behavior of $\mathcal{L}_{\text{CARL}}$ in \textbf{Appendix~\ref{sec:bound_carl}}.
The analysis shows that CARL reduces parameter error through two structural mechanisms, achieving more stable and efficient fallibility representation learning:
\begin{enumerate}[leftmargin=*]
    \item \textbf{Variance isolation:} The variance introduced by sampling $l_i$ is isolated into the $\mathcal{L}_{\text{CL}}$ term, which typically has a smaller weight, thus reducing its contribution to the gradient variance.
    \item \textbf{Negative covariance constraint:} The shared negative set induces a critical negative covariance term, which partially cancels the variance generated by component sampling $c_j$.
\end{enumerate}

\newcolumntype{C}{>{\centering\arraybackslash}X}

\begin{table*}[htbp]
  \centering
  \captionsetup{skip=2pt}
  \scriptsize\scalefont{1.0}
  \renewcommand{\arraystretch}{1.0}
  \caption{\small Comparison results with baseline methods for root cause localization.}
  \resizebox{1.0\textwidth}{!}{%
  \begin{tabularx}{\textwidth}{c l *{8}{C}}
    \toprule
    \textbf{Dataset} & \textbf{Embedding Model} & \textbf{HR@1} & \textbf{HR@3} & \textbf{HR@5} & \textbf{PR@3} & \textbf{PR@5} & \textbf{MAP@3} & \textbf{MAP@5} & \textbf{MRR} \\
    \midrule
    \multirow{14}{*}{\textbf{BGL}} & \multicolumn{9}{>{\columncolor{small_models}}c}{{\textbf{\textit{General Pretrained Small Language Models}}}} \\
      & Glove-300d & 40.00 ± 6.00 & 52.55 ± 5.39 & 56.29 ± 2.86 & 44.32 ± 5.02 & 45.81 ± 3.61 & 42.41 ± 5.46 & 43.65 ± 4.77 & 46.62 ± 5.13 \\
      & BERT-Base & 58.64 ± 10.08 & 73.21 ± 12.35 & 79.83 ± 11.89 & 61.63 ± 5.87 & 65.62 ± 6.31 & 59.65 ± 6.95 & 61.70 ± 6.43 & 67.40 ± 8.81 \\
      & ALBERT-Base & 53.74 ± 0.58 & 61.24 ± 2.68 & 68.07 ± 4.00 & 55.67 ± 1.11 & 58.13 ± 1.40 & 54.72 ± 0.76 & 55.78 ± 0.87 & 61.03 ± 1.07 \\
      & BART-Base & 62.47 ± 5.26 & 78.10 ± 13.27 & 84.35 ± 12.71 & 68.04 ± 8.08 & 69.60 ± 8.63 & 65.38 ± 6.44 & 66.93 ± 7.29 & 72.11 ± 6.95 \\
    \cmidrule(lr){2-10}
      & \multicolumn{9}{>{\columncolor{large_models}}c}{{\textbf{\textit{General Pretrained Large Language Models}}}} \\
      & GTE-Qwen-1.5B-Instruct & 81.24 ± 3.51 & 88.65 ± 3.86 & 90.05 ± 4.01 & 85.13 ± 3.92 & 85.60 ± 3.63 & 83.44 ± 3.78 & 84.22 ± 3.64 & 85.08 ± 3.73 \\
      & Stella\_en\_1.5B\_v5 & 69.46 ± 7.23 & 89.01 ± 5.70 & 91.69 ± 3.49 & 75.78 ± 4.27 & 78.54 ± 4.10 & 73.47 ± 4.68 & 75.28 ± 4.33 & 79.55 ± 5.46 \\
      & E5-mistral-7b-instruct & 78.85 ± 12.42 & 94.14 ± 2.61 & 95.02 ± 2.90 & 78.76 ± 4.52 & 79.29 ± 4.90 & 79.09 ± 6.39 & 79.25 ± 5.64 & 86.52 ± 7.13 \\
      & SFR-Embedding-Mistral & 85.67 ± 4.96 & 92.35 ± 3.90 & 93.38 ± 3.61 & 88.57 ± 4.36 & 90.02 ± 4.00 & 87.16 ± 4.59 & 88.18 ± 4.36 & 89.06 ± 4.39 \\
    \cmidrule(lr){2-10}
      & \multicolumn{9}{>{\columncolor{log_models}}c}{{\textbf{\textit{Log-Specific Models}}}} \\
      & PreLog & 54.16 ± 4.64 & 73.87 ± 9.27 & 80.08 ± 6.26 & 53.30 ± 4.53 & 55.26 ± 4.25 & 53.17 ± 3.82 & 53.83 ± 3.80 & 64.94 ± 4.84 \\
      & BART-Log & 61.73 ± 7.11 & 69.86 ± 10.34 & 74.21 ± 9.15 & 57.18 ± 6.32 & 60.78 ± 6.18 & 58.54 ± 6.28 & 59.06 ± 6.16 & 68.46 ± 6.96 \\
    \rowcolor{\highlightcolor}
    \multicolumn{1}{c}{\cellcolor{white}} & Bifrost (Ours) & \textbf{90.10 ± 1.25} & \textbf{94.67 ± 1.57} & \textbf{95.12 ± 1.45} & \textbf{92.40 ± 1.29} & \textbf{93.38 ± 1.22} & \textbf{91.22 ± 1.23} & \textbf{92.00 ± 1.25} & \textbf{92.43 ± 1.33} \\
    \midrule
    \multirow{14}{*}{\textbf{Thunderbird}} & \multicolumn{9}{>{\columncolor{small_models}}c}{{\textbf{\textit{General Pretrained Small Language Models}}}} \\
      & Glove-300d & 8.07 ± 2.23 & 38.08 ± 5.15 & 54.78 ± 3.76 & 29.80 ± 3.41 & 42.67 ± 3.28 & 19.71 ± 2.73 & 27.68 ± 2.95 & 27.19 ± 2.50 \\
      & BERT-Base & 25.77 ± 8.81 & 36.85 ± 1.49 & 42.44 ± 2.92 & 30.59 ± 3.86 & 37.16 ± 2.60 & 28.31 ± 5.84 & 31.22 ± 4.36 & 34.84 ± 4.61 \\
      & ALBERT-Base & 38.43 ± 8.43 & 52.88 ± 4.37 & 61.96 ± 3.75 & 42.99 ± 5.48 & 50.85 ± 4.55 & 40.20 ± 6.73 & 43.67 ± 5.83 & 48.36 ± 5.96 \\
      & BART-Base & 44.83 ± 5.08 & 58.94 ± 6.45 & 66.67 ± 6.63 & 46.54 ± 5.14 & 54.35 ± 5.13 & 45.61 ± 4.67 & 48.33 ± 4.62 & 55.37 ± 4.26 \\
    \cmidrule(lr){2-10}
      & \multicolumn{9}{>{\columncolor{large_models}}c}{{\textbf{\textit{General Pretrained Large Language Models}}}} \\
      & GTE-Qwen-1.5B-Instruct & 41.98 ± 5.96 & 67.47 ± 6.76 & 84.79 ± 7.48 & 56.96 ± 3.18 & 72.35 ± 5.38 & 49.36 ± 4.00 & 57.15 ± 4.42 & 58.29 ± 5.12 \\
      & Stella\_en\_1.5B\_v5 & 52.65 ± 8.62 & 59.95 ± 9.53 & 69.08 ± 9.82 & 56.98 ± 9.45 & 63.63 ± 9.50 & 54.59 ± 8.95 & 57.58 ± 9.13 & 59.80 ± 8.77 \\
      & E5-mistral-7b-instruct & 42.13 ± 10.22 & 70.16 ± 9.36 & 83.28 ± 5.43 & 57.29 ± 6.14 & 68.35 ± 4.96 & 49.80 ± 7.67 & 56.17 ± 6.69 & 59.08 ± 8.14 \\
      & SFR-Embedding-Mistral & 47.38 ± 6.05 & 77.98 ± 7.70 & 91.33 ± 3.67 & 61.34 ± 5.63 & 72.84 ± 4.84 & 53.97 ± 5.81 & 60.45 ± 5.37 & 64.27 ± 5.54 \\
    \cmidrule(lr){2-10}
      & \multicolumn{9}{>{\columncolor{log_models}}c}{{\textbf{\textit{Log-Specific Models}}}} \\
      & PreLog & 52.93 ± 12.22 & 74.30 ± 8.04 & 77.74 ± 7.71 & 61.66 ± 6.89 & 68.44 ± 7.26 & 57.47 ± 8.85 & 61.29 ± 7.68 & 63.40 ± 9.29 \\
      & BART-Log & 42.79 ± 7.44 & 45.90 ± 4.10 & 51.56 ± 4.43 & 44.17 ± 4.83 & 46.20 ± 4.09 & 43.53 ± 5.75 & 44.38 ± 4.94 & 49.57 ± 4.94 \\
    \rowcolor{\highlightcolor}
    \multicolumn{1}{c}{\cellcolor{white}} & Bifrost (Ours) & \textbf{75.77 ± 5.16} & \textbf{83.85 ± 5.51} & \textbf{86.07 ± 5.51} & \textbf{81.41 ± 5.31} & \textbf{83.18 ± 5.43} & \textbf{78.80 ± 5.29} & \textbf{80.37 ± 5.31} & \textbf{80.32 ± 5.31} \\
    \midrule
    \multirow{14}{*}{\textbf{Platform-X}} & \multicolumn{9}{>{\columncolor{small_models}}c}{{\textbf{\textit{General Pretrained Small Language Models}}}} \\
      & Glove-300d & 52.44 ± 3.17 & 70.24 ± 0.98 & 78.21 ± 1.83 & 56.29 ± 1.69 & 60.36 ± 1.45 & 54.22 ± 2.22 & 56.29 ± 1.55 & 62.64 ± 1.86 \\
      & BERT-Base & 72.75 ± 0.77 & 79.05 ± 0.80 & 79.68 ± 0.70 & 71.43 ± 0.91 & 70.69 ± 0.78 & 72.05 ± 0.81 & 71.56 ± 0.76 & 75.91 ± 0.78 \\
      & ALBERT-Base & 72.41 ± 1.84 & 81.42 ± 0.45 & 82.69 ± 0.21 & 72.70 ± 1.29 & 73.49 ± 1.39 & 72.64 ± 1.54 & 72.90 ± 1.44 & 76.98 ± 1.13 \\
      & BART-Base & 74.80 ± 1.32 & 79.17 ± 1.16 & 80.91 ± 0.93 & 73.52 ± 0.95 & 74.62 ± 0.92 & 74.00 ± 1.06 & 74.11 ± 0.97 & 77.27 ± 1.04 \\
    \cmidrule(lr){2-10}
      & \multicolumn{9}{>{\columncolor{large_models}}c}{{\textbf{\textit{General Pretrained Large Language Models}}}} \\
      & GTE-Qwen-1.5B-Instruct & 83.32 ± 0.37 & 84.91 ± 0.18 & 85.24 ± 0.34 & 80.37 ± 0.35 & 80.97 ± 0.73 & 81.78 ± 0.33 & 81.37 ± 0.36 & 84.19 ± 0.14 \\
      & Stella\_en\_1.5B\_v5 & 83.15 ± 0.20 & 84.68 ± 0.59 & 85.67 ± 0.49 & 80.01 ± 0.80 & 80.35 ± 0.98 & 81.36 ± 0.66 & 80.91 ± 0.73 & 84.10 ± 0.33 \\
      & E5-mistral-7b-instruct & 84.89 ± 0.76 & 86.29 ± 0.11 & 86.33 ± 0.08 & 79.74 ± 0.99 & 78.63 ± 1.05 & 82.13 ± 0.93 & 80.78 ± 0.64 & 85.57 ± 0.45 \\
      & SFR-Embedding-Mistral & 84.69 ± 0.98 & 86.17 ± 0.36 & 86.26 ± 0.28 & 80.61 ± 1.05 & 79.95 ± 1.28 & 82.50 ± 0.64 & 81.49 ± 0.58 & 85.43 ± 0.65 \\
    \cmidrule(lr){2-10}
      & \multicolumn{9}{>{\columncolor{log_models}}c}{{\textbf{\textit{Log-Specific Models}}}} \\
      & PreLog & 69.91 ± 1.61 & 77.20 ± 2.08 & 78.02 ± 2.06 & 71.95 ± 1.83 & 72.41 ± 2.04 & 71.11 ± 1.67 & 71.56 ± 1.78 & 73.62 ± 1.86 \\
      & BART-Log & 57.41 ± 1.13 & 77.98 ± 0.54 & 80.35 ± 0.44 & 62.80 ± 0.91 & 64.57 ± 1.08 & 60.76 ± 0.89 & 62.13 ± 0.84 & 69.11 ± 0.74 \\
    \rowcolor{\highlightcolor}
    \multicolumn{1}{c}{\cellcolor{white}} & Bifrost (Ours) & \textbf{87.98 ± 0.96} & \textbf{92.35 ± 0.86} & \textbf{93.08 ± 0.94} & \textbf{88.36 ± 0.95} & \textbf{88.40 ± 1.42} & \textbf{88.33 ± 0.85} & \textbf{88.35 ± 1.00} & \textbf{90.14 ± 0.93} \\
    \bottomrule
  \end{tabularx}%
  }
  \label{tab:leaderboard_rca}
  \vspace{-0.4cm}
\end{table*}

\begin{table*}[htbp]
  \centering
  \captionsetup{skip=2pt}
  \scriptsize\scalefont{1.0}
  \caption{\small Comparison results with baseline methods for anomaly detection.}
  \resizebox{1.0\textwidth}{!}{%
  \setlength{\tabcolsep}{4pt}
  \begin{tabular}{l c ccc ccc ccc ccc}
    \toprule
    \multirow{2}{*}{\textbf{Embedding Model}} & \multirow{2}{*}{\textbf{Parameters}} & \multicolumn{3}{c}{\textbf{BGL}} & \multicolumn{3}{c}{\textbf{Hadoop}} & \multicolumn{3}{c}{\textbf{Thunderbird}} & \multicolumn{3}{c}{\textbf{Platform-X}} \\
    \cmidrule(lr){3-5} \cmidrule(lr){6-8} \cmidrule(lr){9-11} \cmidrule(lr){12-14}
    & & \textbf{\makecell{Precision}} & \textbf{\makecell{Recall}} & \textbf{\makecell{F1-Score}} & \textbf{\makecell{Precision}} & \textbf{\makecell{Recall}} & \textbf{\makecell{F1-Score}} & \textbf{\makecell{Precision}} & \textbf{\makecell{Recall}} & \textbf{\makecell{F1-Score}} & \textbf{\makecell{Precision}} & \textbf{\makecell{Recall}} & \textbf{\makecell{F1-Score}} \\
    \midrule
    \rowcolor{small_models} \multicolumn{14}{c}{ \textbf{\textit{General Pretrained Small Language Models}} } \\
    Glove-300d & 120M  & 88.41 ± 3.54 & 92.06 ± 0.33 & 90.16 ± 1.84 & 85.11 ± 4.31 & 94.82 ± 0.43 & 89.65 ± 2.51 & 84.67 ± 4.22 & 70.15 ± 13.98 & 76.13 ± 10.69 & 88.20 ± 2.73 & 86.73 ± 2.85 & 87.37 ± 0.41 \\
    BERT-Base & 110M  & 93.90 ± 1.42 & 89.96 ± 1.33 & 91.89 ± 1.36 & 89.04 ± 9.11 & 94.92 ± 0.37 & 91.65 ± 5.38 & 91.50 ± 2.08 & 81.19 ± 0.37 & 86.03 ± 0.92 & 88.29 ± 2.07 & 92.78 ± 0.16 & 90.47 ± 1.07 \\
    ALBERT-Base & 12M   & 94.60 ± 0.61 & 92.34 ± 0.08 & 93.46 ± 0.29 & 50.87 ± 3.33 & 94.65 ± 0.14 & 66.11 ± 2.74 & 97.25 ± 0.78 & 77.88 ± 2.38 & 86.46 ± 1.26 & 90.28 ± 2.20 & 87.26 ± 1.04 & 88.71 ± 0.73 \\
    BART-Base & 140M  & 90.53 ± 4.11 & 91.96 ± 1.22 & 91.18 ± 1.88 & 70.18 ± 8.94 & 94.56 ± 0.39 & 80.27 ± 5.49 & 92.60 ± 3.82 & 79.97 ± 1.93 & 85.74 ± 1.26 & 81.37 ± 2.39 & 84.44 ± 2.35 & 82.83 ± 1.37 \\
    \midrule
    \rowcolor{large_models} \multicolumn{14}{c}{ \textbf{\textit{General Pretrained Large Language Models} }} \\
    GTE-Qwen-1.5B-Instruct & 1800M & 91.85 ± 3.52 & 93.47 ± 1.45 & 92.63 ± 2.36 & 60.40 ± 12.50 & 94.94 ± 0.30 & 73.13 ± 8.53 & 96.94 ± 1.03 & 80.79 ± 0.77 & 88.13 ± 0.67 & 98.72 ± 0.36 & 90.54 ± 0.22 & 94.45 ± 0.16 \\
    Stella\_en\_1.5B\_v5 & 1500M & 90.69 ± 2.64 & 90.73 ± 2.92 & 90.71 ± 2.68 & 63.99 ± 9.16 & 95.15 ± 0.29 & 76.14 ± 6.52 & 96.53 ± 1.97 & 83.20 ± 1.45 & 89.35 ± 0.80 & 98.92 ± 0.72 & 90.56 ± 0.17 & 94.55 ± 0.28 \\
    E5-mistral-7b-instruct & 7200M & 98.45 ± 0.94 & 94.16 ± 0.90 & 96.25 ± 0.53 & 78.84 ± 10.00 & 95.32 ± 0.52 & 85.93 ± 6.02 & 98.75 ± 0.42 & 84.18 ± 2.39 & 90.86 ± 1.35 & 99.13 ± 0.39 & 91.05 ± 0.20 & 94.92 ± 0.12 \\
    SFR-Embedding-Mistral & 7200M & 98.65 ± 0.34 & 95.72 ± 2.16 & 97.15 ± 0.99 & 83.12 ± 13.88 & 95.30 ± 0.66 & 88.13 ± 8.42 & 98.59 ± 0.37 & 81.27 ± 9.32 & 88.81 ± 5.65 & 99.08 ± 0.61 & 90.95 ± 0.25 & 94.84 ± 0.24 \\
    \midrule
    \rowcolor{log_models} \multicolumn{14}{c}{ \textbf{\textit{Log-Specific Models}} } \\
    PreLog & 140M  & 95.94 ± 2.31 & 89.38 ± 1.20 & 92.52 ± 0.91 & 72.72 ± 12.24 & 96.52 ± 0.13 & 82.38 ± 7.82 & 79.17 ± 11.60 & 75.98 ± 7.43 & 77.14 ± 8.11 & 88.34 ± 5.09 & 82.78 ± 5.90 & 85.12 ± 1.50 \\
    BART-Log & 140M  & 94.26 ± 2.92 & 91.46 ± 1.73 & 92.84 ± 2.07 & 79.55 ± 9.93 & 94.38 ± 0.34 & 86.33 ± 6.48 & 91.06 ± 2.22 & 74.74 ± 1.13 & 82.10 ± 0.91 & 95.76 ± 1.38 & 86.60 ± 0.73 & 90.95 ± 0.72 \\
    \rowcolor{\highlightcolor}
    Bifrost (Ours) & 140M & \textbf{98.81 ± 0.67} & \textbf{98.34 ± 0.53} & \textbf{98.57 ± 0.52} & \textbf{97.56 ± 0.81} & \textbf{96.14 ± 0.60} & \textbf{96.84 ± 0.22} & \textbf{97.50 ± 2.09} & \textbf{90.78 ± 2.13} & \textbf{94.01 ± 1.72} & \textbf{96.33 ± 0.76} & \textbf{97.25 ± 0.95} & \textbf{96.78 ± 0.21} \\
    \bottomrule
  \end{tabular}%
  }
  \label{tab:leaderboard_ad}
  \vspace{-0.4cm}
\end{table*}

\begin{table}[htbp]
  \centering
  \captionsetup{skip=2pt}
  \caption{\small Evaluation results on fault identification task.}
  \label{tab:leaderboard_fi}

  \resizebox{0.85\linewidth}{!}{%
    \begin{tabular}{lccc}
      \toprule
      \textbf{Embedding Model} &
      \textbf{\makecell{Macro-Precision}} &
      \textbf{\makecell{Macro-Recall}} &
      \textbf{\makecell{Macro-F1-Score}} \\
      \midrule
      \rowcolor{small_models} \multicolumn{4}{c}{\textbf{\textit{General Pretrained Small Language Models}} } \\
      Glove-300d & 80.95 ± 0.50 & 77.29 ± 1.95 & 77.26 ± 1.13 \\
      BERT-Base & 77.68 ± 2.65 & 72.11 ± 3.28 & 72.92 ± 3.15 \\
      ALBERT-Base & 73.62 ± 8.14 & 71.44 ± 4.59 & 70.76 ± 6.15 \\
      BART-Base & 47.47 ± 5.94 & 66.15 ± 1.92 & 54.86 ± 4.27 \\
      \midrule
      \rowcolor{large_models} \multicolumn{4}{c}{\textbf{\textit{General Pretrained Large Language Models}}} \\
      GTE-Qwen-1.5B-Instruct & 97.66 ± 0.65 & 88.02 ± 0.63 & 92.45 ± 0.56 \\
      Stella\_en\_1.5B\_v5 & 96.77 ± 1.69 & 86.40 ± 3.11 & 90.76 ± 2.94 \\
      E5-mistral-7b-instruct & 97.80 ± 1.27 & 88.25 ± 1.63 & 92.43 ± 1.76 \\
      SFR-Embedding-Mistral & 97.78 ± 0.56 & 88.55 ± 0.61 & 92.73 ± 0.64 \\
      \midrule
      \rowcolor{log_models} \multicolumn{4}{c}{\textbf{\textit{Log-Specific Models}}} \\
      PreLog & 51.00 ± 8.46 & 65.02 ± 1.01 & 56.38 ± 4.74 \\
      BART-Log & 39.51 ± 4.86 & 80.31 ± 2.82 & 52.96 ± 3.85 \\
      \rowcolor{\highlightcolor}
      Bifrost (Ours) & \textbf{94.84 ± 0.44} & \textbf{92.81 ± 0.53} & \textbf{93.35 ± 0.43} \\
      \bottomrule
    \end{tabular}%
  }

  \vspace{-0.4cm}
\end{table}

\section{Experimental Evaluation}
In this section, we evaluate our proposed method by answering the following research questions:
\begin{itemize}[leftmargin=*]
    \item \textbf{RQ1:} How effective is fallibility representation in the log-based fault diagnosis?
    \item \textbf{RQ2:} What are the contributions of different training strategies?
    \item \textbf{RQ3:} Does \textsc{Bifrost} over-rely on the injected failure lexicon in fallibility learning?
    \item \textbf{RQ4:} How capable is \textsc{Bifrost} of encoding the fallibility representation?
\end{itemize}

\subsection{Experimental Setup}
\label{sec:appendix_exp}

\subsubsection{\textbf{Datasets}}
We evaluate \textsc{Bifrost} on four log datasets: three public benchmarks (BGL, Thunderbird, and Hadoop) and one industrial dataset from a production MLaaS platform (Platform-X) \cite{zhu2023loghub, oliner2007supercomputers}. 
BGL and Thunderbird come from the Blue Gene/L supercomputer (128K processors) and the Thunderbird supercomputer (9,024 processors), respectively \cite{oliner2007supercomputers}; Hadoop comes from a five-node HDFS cluster with 46 CPU cores \cite{zhu2023loghub}. 
Platform-X operates over 6,000 heterogeneous GPUs to run ML jobs. We collect its system logs over a 7-day workload, and SREs manually label failures into six component categories: hardware, kernel, disk, memory, CPU, and GPU. 
Following prior work \cite{le2021log, le2022log}, we use the earliest 10 million log messages from Thunderbird due to its scale.

\subsubsection{\textbf{Baselines}}
To evaluate \textsc{Bifrost}, we compare against (i) general PLMs in the $\sim$100M-parameter regime \cite{devlin2019bert, lewis2020bart, lanalbert, pennington2014glove}, (ii) larger general PLMs with $>$1B parameters \cite{li2023towards, zhang2025jasperstelladistillationsota, wang2023improving, SFRAIResearch2024}, and (iii) log-specific models \cite{le2024prelog}. 
Additionally, we introduce a baseline, \textsc{BART-Log}, which shares the same BART-Base backbone and is further trained on the same training-split logs as \textsc{Bifrost}, but with the original pretraining objective. This design isolates the effect of additional training on target-system logs, enabling fair attribution of gains to the proposed fallibility-aware objectives.

\subsubsection{\textbf{Evaluation Task}}
We evaluate \textsc{Bifrost} on three fault diagnosis sub-tasks: AD, RCL, and FI.
For AD, we use BGL, Hadoop, Thunderbird, and Platform-X, and report Precision, Recall, and F1 \cite{du2017deeplog, le2022log}.
For RCL, we use BGL, Thunderbird, and Platform-X because Hadoop lacks root-cause labels/logs, and report HR@k, NDCG@k, MAP@k, and MRR with $k \in \{3,5\}$ \cite{wittkopp2024logrca, he2025united}.
For FI, we use Platform-X only because public datasets do not provide fault-type labels, and report Macro-Precision, Macro-Recall, and Macro-F1 \cite{zhang2021cloudrca, duan2025famos, sui2023logkg}.

\subsubsection{\textbf{Implementation Details}}
\textbf{Fault Diagnosis Evaluation.}
For AD/RCL/FI, we use LogRobust \cite{zhang2019robust}, LogRCA \cite{wittkopp2024logrca}, and CloudLog \cite{zhang2021cloudrca} as backbones, respectively.
Following prior work \cite{meng2019loganomaly, le2022log}, we apply Drain parsing \cite{he2017drain}, group logs with a sliding window of 20, and split data into train/val/test with a 6:1:3 ratio.
All results are averaged over five runs (mean$\pm$std).
Experiments are conducted on a Linux server with an NVIDIA H20 GPU.
For all baselines, we use their public implementations and default settings \cite{le2022log}.

\textbf{\textsc{Bifrost} Training.}
\textsc{Bifrost} is built on BART-Base \cite{lewis2020bart}.
We set $(\lambda_1,\lambda_2,\lambda_3,\lambda_4)=(1,0.5,1,1)$ for the joint loss, and set the system failure corpus size to 10 and the system component corpus size to 8.
To prevent data leakage, fallibility representation learning and fault lexicon construction are performed using only normal logs from the training set, the validation set is used solely for hyperparameter tuning, and the test set is never involved in training.

\begin{table*}[htbp]
  \centering
  \captionsetup{skip=2pt}
  \scriptsize\scalefont{1.0}
  \renewcommand{\arraystretch}{1.0}
  \caption{\small Ablation results on the root cause localization task.}
  \label{tab:addlabel}
  \begin{tabularx}{1.0\textwidth}{c cccc CCCCCCCC}
    \toprule
    \multirow{2}{*}{\textbf{Dataset}} & \multicolumn{4}{c}{\textbf{Learning Objective}} & \multicolumn{8}{c}{\textbf{Metrics}} \\
    \cmidrule(lr){2-5} \cmidrule(lr){6-13}
    & \textbf{CARL} & \textbf{EFP} & \textbf{AED} & \textbf{SCP} & \textbf{HR@1} & \textbf{HR@3} & \textbf{HR@5} & \textbf{PR@3} & \textbf{PR@5} & \textbf{MAP@3} & \textbf{MAP@5} & \textbf{MRR} \\
    \midrule
    \multirow{8}{*}{\textbf{BGL}} 
    & \Checkmark & \Checkmark & \XSolidBrush & \XSolidBrush & 82.85 ± 3.59 & 88.86 ± 3.74 & 89.69 ± 4.00 & 85.55 ± 3.47 & 86.23 ± 3.16 & 84.23 ± 3.44 & 85.02 ± 3.38 & 85.81 ± 3.66 \\
    & \Checkmark & \XSolidBrush & \Checkmark & \XSolidBrush & 81.24 ± 3.51 & 88.65 ± 3.86 & 90.05 ± 4.01 & 85.13 ± 3.92 & 85.60 ± 3.63 & 83.44 ± 3.78 & 84.22 ± 3.64 & 85.08 ± 3.73 \\
    & \Checkmark & \XSolidBrush & \XSolidBrush & \Checkmark & 83.05 ± 3.99 & 89.05 ± 3.41 & 90.54 ± 3.25 & 81.21 ± 9.94 & 82.48 ± 10.9 & 81.70 ± 7.39 & 81.89 ± 8.74 & 86.20 ± 3.54 \\
    & \Checkmark & \XSolidBrush & \Checkmark & \Checkmark & 85.98 ± 3.40 & 91.29 ± 2.99 & 92.08 ± 2.74 & 88.39 ± 3.14 & 89.34 ± 3.02 & 87.24 ± 3.22 & 88.01 ± 3.15 & 88.80 ± 3.09 \\
    & \Checkmark & \Checkmark & \XSolidBrush & \Checkmark & 86.52 ± 1.98 & 91.86 ± 1.13 & 92.42 ± 1.05 & 89.34 ± 1.15 & 90.22 ± 1.06 & 88.05 ± 1.46 & 88.85 ± 1.30 & 89.35 ± 1.43 \\
    & \Checkmark & \Checkmark & \Checkmark & \XSolidBrush & 87.86 ± 1.11 & 92.57 ± 1.33 & 93.31 ± 1.52 & 90.01 ± 1.11 & 91.03 ± 1.26 & 88.96 ± 1.12 & 89.70 ± 1.14 & 90.44 ± 1.18 \\
    \cmidrule(lr){2-13}
    & \XSolidBrush & \Checkmark & \Checkmark & \Checkmark & 83.92 ± 1.79 & 90.27 ± 1.90 & 91.39 ± 1.76 & 87.55 ± 1.77 & 88.91 ± 1.72 & 85.80 ± 1.71 & 86.95 ± 1.71 & 87.32 ± 1.56 \\
    \cmidrule(lr){2-13}
    \rowcolor{\highlightcolor}
    \cellcolor{white} & \Checkmark & \Checkmark & \Checkmark & \Checkmark & \textbf{90.10 ± 1.25} & \textbf{94.67 ± 1.57} & \textbf{95.12 ± 1.45} & \textbf{92.40 ± 1.29} & \textbf{93.38 ± 1.22} & \textbf{91.22 ± 1.23} & \textbf{92.00 ± 1.25} & \textbf{92.43 ± 1.33} \\
    \midrule
    \multirow{8}{*}{\textbf{Thunderbird}} 
    & \Checkmark & \Checkmark & \XSolidBrush & \XSolidBrush & 66.79 ± 7.61 & 76.08 ± 6.58 & 77.70 ± 6.71 & 73.02 ± 7.01 & 75.14 ± 6.75 & 70.09 ± 7.24 & 71.94 ± 7.08 & 71.80 ± 7.14 \\
    & \Checkmark & \XSolidBrush & \Checkmark & \XSolidBrush & 67.11 ± 3.77 & 74.66 ± 4.02 & 76.75 ± 3.70 & 72.21 ± 4.08 & 73.80 ± 3.95 & 69.88 ± 4.03 & 71.30 ± 4.01 & 71.21 ± 3.80 \\
    & \Checkmark & \XSolidBrush & \XSolidBrush & \Checkmark & 65.39 ± 5.92 & 75.45 ± 7.08 & 78.28 ± 7.38 & 72.13 ± 6.72 & 74.74 ± 7.09 & 68.86 ± 6.34 & 70.98 ± 6.61 & 70.87 ± 6.34 \\
    & \Checkmark & \XSolidBrush & \Checkmark & \Checkmark & 70.31 ± 4.06 & 79.23 ± 3.62 & 81.69 ± 3.94 & 76.00 ± 3.86 & 78.29 ± 3.83 & 73.26 ± 3.93 & 75.06 ± 3.82 & 75.23 ± 3.72 \\
    & \Checkmark & \Checkmark & \XSolidBrush & \Checkmark & 68.99 ± 9.46 & 78.49 ± 8.54 & 81.13 ± 8.58 & 74.94 ± 8.45 & 77.49 ± 8.46 & 72.01 ± 8.83 & 73.96 ± 8.68 & 74.16 ± 8.89 \\
    & \Checkmark & \Checkmark & \Checkmark & \XSolidBrush & 71.70 ± 4.35 & 81.49 ± 6.27 & 84.08 ± 6.60 & 78.49 ± 5.44 & 80.83 ± 6.04 & 75.37 ± 4.90 & 77.34 ± 5.32 & 76.98 ± 5.08 \\
    \cmidrule(lr){2-13}
    & \XSolidBrush & \Checkmark & \Checkmark & \Checkmark & 70.67 ± 2.84 & 83.65 ± 3.86 & 85.72 ± 3.58 & 78.81 ± 2.91 & 81.67 ± 3.25 & 74.90 ± 2.82 & 77.36 ± 2.88 & 77.34 ± 2.98 \\
    \cmidrule(lr){2-13}
    \rowcolor{\highlightcolor}
    \cellcolor{white} & \Checkmark & \Checkmark & \Checkmark & \Checkmark & \textbf{75.77 ± 5.16} & \textbf{83.85 ± 5.51} & \textbf{86.07 ± 5.51} & \textbf{81.41 ± 5.31} & \textbf{83.18 ± 5.43} & \textbf{78.80 ± 5.29} & \textbf{80.37 ± 5.31} & \textbf{80.32 ± 5.31} \\
    \midrule
    \multirow{8}{*}{\textbf{Platform-X}} 
    & \Checkmark & \Checkmark & \XSolidBrush & \XSolidBrush & 82.43 ± 1.58 & 88.55 ± 1.86 & 90.49 ± 1.93 & 81.57 ± 1.90 & 80.45 ± 2.17 & 82.18 ± 1.73 & 81.60 ± 1.85 & 85.62 ± 1.69 \\
    & \Checkmark & \XSolidBrush & \Checkmark & \XSolidBrush & 82.52 ± 0.71 & 89.05 ± 0.97 & 91.45 ± 1.13 & 79.43 ± 2.04 & 77.73 ± 2.31 & 81.19 ± 1.38 & 79.94 ± 1.75 & 86.12 ± 0.76 \\
    & \Checkmark & \XSolidBrush & \XSolidBrush & \Checkmark & 81.98 ± 1.15 & 87.79 ± 2.05 & 89.66 ± 2.57 & 80.38 ± 1.99 & 79.09 ± 2.50 & 81.30 ± 1.22 & 80.51 ± 1.55 & 85.09 ± 1.65 \\
    & \Checkmark & \XSolidBrush & \Checkmark & \Checkmark & 85.61 ± 1.91 & 90.89 ± 1.18 & 91.95 ± 0.93 & 85.57 ± 2.55 & 84.57 ± 3.74 & 85.81 ± 2.18 & 85.39 ± 2.68 & 88.27 ± 1.45 \\
    & \Checkmark & \Checkmark & \XSolidBrush & \Checkmark & 84.07 ± 1.60 & 89.22 ± 1.47 & 90.36 ± 1.33 & 83.94 ± 2.77 & 83.04 ± 3.87 & 84.16 ± 2.30 & 83.78 ± 2.82 & 86.68 ± 1.41 \\
    & \Checkmark & \Checkmark & \Checkmark & \XSolidBrush & 86.64 ± 2.52 & 92.32 ± 1.49 & 93.78 ± 1.04 & 84.95 ± 2.38 & 84.16 ± 2.70 & 85.93 ± 2.36 & 85.27 ± 2.45 & 89.60 ± 1.90 \\
    \cmidrule(lr){2-13}
    & \XSolidBrush & \Checkmark & \Checkmark & \Checkmark & 86.55 ± 0.24 & 91.98 ± 0.65 & 92.86 ± 0.64 & 85.60 ± 0.68 & 84.64 ± 1.81 & 86.29 ± 0.44 & 85.74 ± 0.86 & 89.27 ± 0.31 \\
    \cmidrule(lr){2-13}
    \rowcolor{\highlightcolor}
    \cellcolor{white} & \Checkmark & \Checkmark & \Checkmark & \Checkmark & \textbf{87.98 ± 0.96} & \textbf{92.35 ± 0.86} & \textbf{93.08 ± 0.94} & \textbf{88.36 ± 0.95} & \textbf{88.40 ± 1.42} & \textbf{88.33 ± 0.85} & \textbf{88.35 ± 1.00} & \textbf{90.14 ± 0.93} \\
    \bottomrule
  \end{tabularx}
  \label{tab:ablation_rca}
\end{table*}

\begin{table*}[htbp]
  \centering
  \captionsetup{skip=2pt}
  \scriptsize\scalefont{1.0}
  \caption{\small Ablation results on the anomaly detection task.}
  \resizebox{1.0\textwidth}{!}{%
  \setlength{\tabcolsep}{5pt}
  \begin{tabular}{cccc ccc ccc ccc ccc}
    \toprule
    \multicolumn{4}{c}{\textbf{Learning Objective}} & \multicolumn{3}{c}{\textbf{BGL}} & \multicolumn{3}{c}{\textbf{Hadoop}} & \multicolumn{3}{c}{\textbf{Thunderbird}} & \multicolumn{3}{c}{\textbf{Platform-X}} \\
    \cmidrule(lr){1-4} \cmidrule(lr){5-7} \cmidrule(lr){8-10} \cmidrule(lr){11-13} \cmidrule(lr){14-16}
    \textbf{CARL} & \textbf{EFP} & \textbf{AED} & \textbf{SCP} & \textbf{Precision} & \textbf{Recall} & \textbf{F1-Score} & \textbf{Precision} & \textbf{Recall} & \textbf{F1-Score} & \textbf{Precision} & \textbf{Recall} & \textbf{F1-Score} & \textbf{Precision} & \textbf{Recall} & \textbf{F1-Score} \\
    \midrule
    \Checkmark & \Checkmark & \XSolidBrush & \XSolidBrush & 98.45 ± 0.52 & 92.52 ± 1.86 & 95.39 ± 1.12 & 95.73 ± 1.43 & 95.55 ± 0.55 & 95.63 ± 0.78 & 96.36 ± 1.09 & 86.31 ± 4.06 & 91.03 ± 2.64 & 92.35 ± 2.09 & 94.46 ± 0.75 & 93.38 ± 1.19 \\
    \Checkmark & \XSolidBrush & \Checkmark & \XSolidBrush & 98.11 ± 0.83 & 92.95 ± 1.59 & 95.45 ± 0.93 & 96.38 ± 1.69 & 95.49 ± 0.44 & 95.93 ± 0.83 & 97.89 ± 1.34 & 82.83 ± 4.77 & 89.63 ± 2.30 & 94.33 ± 1.77 & 89.99 ± 0.68 & 92.10 ± 0.58 \\
    \Checkmark & \XSolidBrush & \XSolidBrush & \Checkmark & 98.10 ± 1.70 & 88.49 ± 1.18 & 93.03 ± 0.36 & 82.29 ± 5.84 & 97.34 ± 0.24 & 89.07 ± 3.44 & 96.15 ± 0.97 & 80.95 ± 0.50 & 87.89 ± 0.63 & 96.15 ± 1.24 & 89.54 ± 0.51 & 92.72 ± 0.48 \\
    \Checkmark & \XSolidBrush & \Checkmark & \Checkmark & 98.66 ± 0.44 & 93.53 ± 2.27 & 96.01 ± 1.06 & 96.44 ± 0.54 & 96.00 ± 0.77 & 96.22 ± 0.40 & 96.45 ± 1.01 & 85.77 ± 2.45 & 90.78 ± 1.57 & 94.20 ± 2.50 & 97.23 ± 0.16 & 95.68 ± 1.27 \\
    \Checkmark & \Checkmark & \XSolidBrush & \Checkmark & 98.70 ± 0.73 & 96.06 ± 2.08 & 97.35 ± 0.99 & 97.66 ± 0.70 & 95.87 ± 0.65 & 96.75 ± 0.30 & 94.77 ± 3.62 & 89.11 ± 2.80 & 91.78 ± 1.90 & 94.54 ± 1.14 & 95.92 ± 0.62 & 95.22 ± 0.35 \\
    \Checkmark & \Checkmark & \Checkmark & \XSolidBrush & 99.15 ± 0.43 & 97.39 ± 1.53 & 98.25 ± 0.64 & 97.51 ± 0.98 & 95.72 ± 0.76 & 96.60 ± 0.49 & 96.84 ± 1.43 & 89.01 ± 2.23 & 92.74 ± 1.43 & 92.67 ± 2.06 & 94.95 ± 0.65 & 93.78 ± 0.91 \\
    \midrule
    \XSolidBrush & \Checkmark & \Checkmark & \Checkmark & 98.84 ± 0.71 & 95.37 ± 3.25 & 97.04 ± 1.76 & 96.33 ± 0.26 & 94.34 ± 2.58 & 95.14 ± 1.58 & 95.18 ± 1.69 & 84.60 ± 2.18 & 89.55 ± 1.27 & 93.23 ± 1.55 & 96.01 ± 0.73 & 94.59 ± 0.69 \\
    \midrule
    \rowcolor{\highlightcolor}
    \Checkmark & \Checkmark & \Checkmark & \Checkmark & \textbf{98.81 ± 0.67} & \textbf{98.34 ± 0.53} & \textbf{98.57 ± 0.52} & \textbf{97.56 ± 0.81} & \textbf{96.14 ± 0.60} & \textbf{96.84 ± 0.22} & \textbf{97.50 ± 2.09} & \textbf{90.78 ± 2.13} & \textbf{94.01 ± 1.72} & \textbf{96.33 ± 0.76} & \textbf{97.25 ± 0.95} & \textbf{96.78 ± 0.21} \\
    \bottomrule
    \end{tabular}%
  }
  \label{tab:ablation_ad}
\end{table*}

\begin{table}[htbp]
  \centering
  \captionsetup{skip=2pt}
  \caption{\small Ablation results on the fault identification task.}
  \label{tab:ablation_fi}

  \resizebox{1.0\linewidth}{!}{%
    \begin{tabular}{ccccccc}
      \toprule
      \multicolumn{4}{c}{\textbf{Learning Objective}} & \multicolumn{3}{c}{\textbf{Platform-X}} \\
      \cmidrule(r){1-4} \cmidrule(l){5-7}
      \textbf{CARL} & \textbf{EFP} & \textbf{AED} & \textbf{SCP} &
      \textbf{\makecell{Macro-Precision}} & \textbf{\makecell{Macro-Recall}} & \textbf{\makecell{Macro-F1}} \\
      \midrule
      \Checkmark & \Checkmark & \XSolidBrush & \XSolidBrush & 81.63 ± 1.25 & 78.62 ± 1.19 & 77.89 ± 1.09 \\
      \Checkmark & \XSolidBrush & \Checkmark & \XSolidBrush & 82.57 ± 1.66 & 78.93 ± 1.91 & 77.82 ± 1.79 \\
      \Checkmark & \XSolidBrush & \XSolidBrush & \Checkmark & 98.17 ± 0.46 & 91.80 ± 1.89 & 94.48 ± 0.89 \\
      \Checkmark & \XSolidBrush & \Checkmark & \Checkmark & 93.00 ± 0.82 & 90.67 ± 1.42 & 90.80 ± 1.65 \\
      \Checkmark & \Checkmark & \XSolidBrush & \Checkmark & 93.06 ± 0.53 & 91.20 ± 0.37 & 91.52 ± 0.49 \\
      \Checkmark & \Checkmark & \Checkmark & \XSolidBrush & 83.18 ± 1.34 & 81.93 ± 1.73 & 80.90 ± 1.79 \\
      \midrule
      \XSolidBrush & \Checkmark & \Checkmark & \Checkmark & 86.95 ± 2.38 & 85.68 ± 1.97 & 85.54 ± 2.17 \\
      \midrule
      \rowcolor{\highlightcolor}
      \Checkmark & \Checkmark & \Checkmark & \Checkmark & \textbf{94.84 ± 0.44} & \textbf{92.81 ± 0.53} & \textbf{93.35 ± 0.43} \\
      \bottomrule
    \end{tabular}%
  }

  \vspace{-0.4cm}
\end{table}

\subsection{RQ1: Effectiveness Evaluation}
Tables \ref{tab:leaderboard_rca}, \ref{tab:leaderboard_ad}, and \ref{tab:leaderboard_fi} report the results on AD, RCL, and FI, respectively. 
Overall, existing small models, large models and log-specific models do not consistently achieve the best performance across these tasks, suggesting that fault diagnosis requires representations better aligned with failure-related log structure. 
In contrast, \textsc{Bifrost} consistently achieves the best results on all three tasks, improving F1 on AD across all four datasets, outperforming all baselines on all RCL ranking metrics, and achieving the highest Macro-F1 on FI.
To further clarify where these gains come from, we compare \textsc{Bifrost} with \textsc{BART-Log}, which uses the same BART-Base backbone and the same training-split logs, but is trained with the original pretraining objective. Although additional training on target-system logs is helpful, \textsc{Bifrost} still consistently outperforms \textsc{BART-Log} across AD, RCL, and FI. This shows that the improvement is not merely due to additional training on target-system logs, but to the proposed fallibility-aware learning objectives, which better capture fault-related execution flow, event semantics, and system components.

\subsection{RQ2: Ablation Study}
This section presents an ablation study to assess the three learning objectives and the CARL strategy in \textsc{Bifrost}. The results are reported in Tables \ref{tab:ablation_rca}, \ref{tab:ablation_ad}, and \ref{tab:ablation_fi}. Overall, every ablated variant degrades performance, demonstrating the effectiveness of these designs. For the three learning objectives, removing a single objective reduces performance by an average of 1.45\%, 2.66\%, and 5.61\% across the three tasks, while removing multiple objectives causes larger drops of 3.95\%, 6.03\%, and 9.95\%. These results suggest that the objectives are complementary and help \textsc{Bifrost} learn fallibility-aware log representations. Removing CARL also leads to clear performance losses of 2.47\%, 2.44\%, and 7.81\%, supporting its role in mitigating fallibility jitter and improving the quality of learned log representations.

\subsection{RQ3: Lexicon Dependence Analysis}
\begin{figure}[t]
\captionsetup{skip=2pt}
\centerline{
\includegraphics[width=1.0\linewidth]{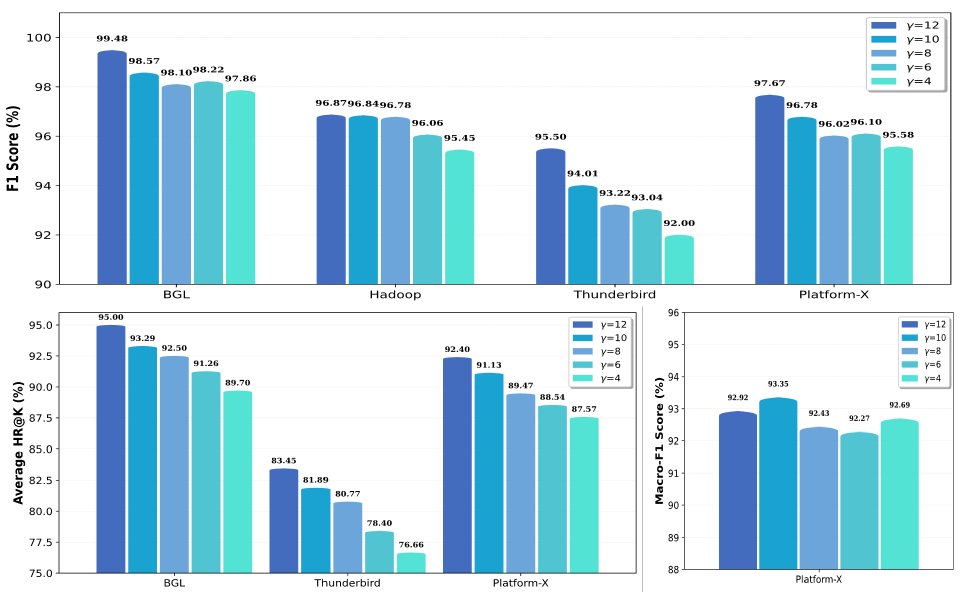}
}
\caption{\small Lexicon dependence of \textsc{Bifrost} with respect to the size of the system failure corpus used for injected failure phrases.}
\Description{Lexicon dependence of \textsc{Bifrost} with respect to the size of the system failure corpus used for injected failure phrases.}
\label{fig:sensity_analysis}
\vspace{-0.4cm}
\end{figure}

We vary the size $K$ of the system failure corpus, which provides failure phrases for EFP and AED, while keeping other settings unchanged. This analysis examines whether the gains of \textsc{Bifrost} mainly come from fitting injected fault phrases. Figure \ref{fig:sensity_analysis} shows the results.
Overall, \textsc{Bifrost} remains strong over a wide range of $K$. AD varies by no more than 2.16\% across datasets, and FI stays within 1.08\% on Platform-X. RCL is more sensitive, with an average span of 5.64\%, suggesting that root-cause ranking benefits more from richer failure cues.
Performance saturates quickly as $K$ increases. Increasing $K$ beyond the default setting yields marginal gains, while reducing $K$ from 10 to 4 causes moderate degradation: AD drops by 0.71\%--2.01\% across datasets, and RCL drops by 3.56\%--5.23\%. These results suggest that \textsc{Bifrost} does not mainly rely on scaling up injected phrases. Since the same failure corpus is shared across systems, the limited variation further suggests that these phrases act as transferable failure cues rather than system-specific lexical shortcuts.

\subsection{RQ4: Representation Analysis}
\begin{figure}[htbp]
    \centering
    \begin{subfigure}[b]{0.48\linewidth}
        \centering
        \includegraphics[width=\linewidth]{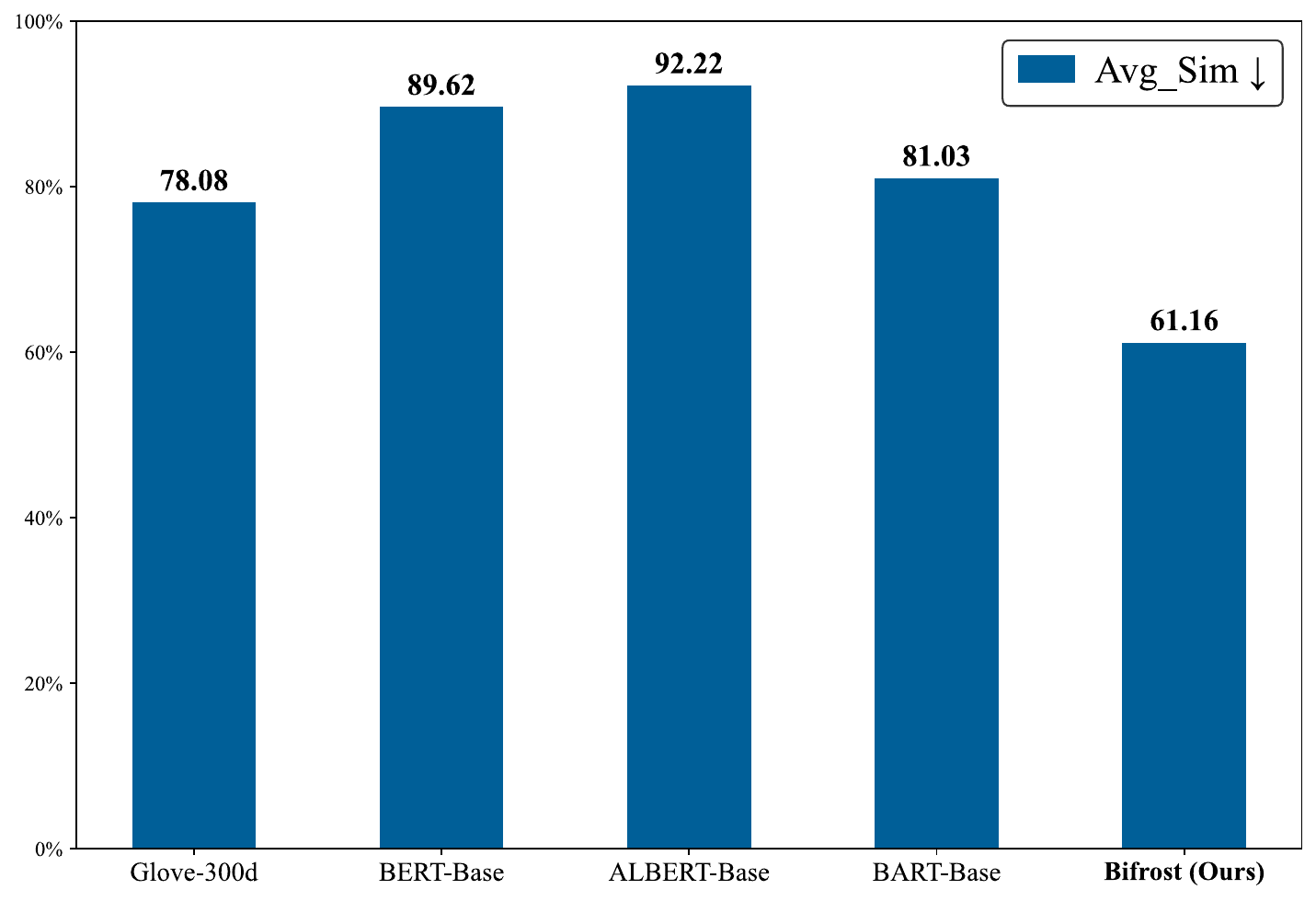}
        \caption{\footnotesize Avg\_Sim as Metric}
        \label{fig:empirical_ad_exp1}
    \end{subfigure}
    \hspace{0.02\linewidth}
    \begin{subfigure}[b]{0.48\linewidth}
        \centering
        \includegraphics[width=\linewidth]{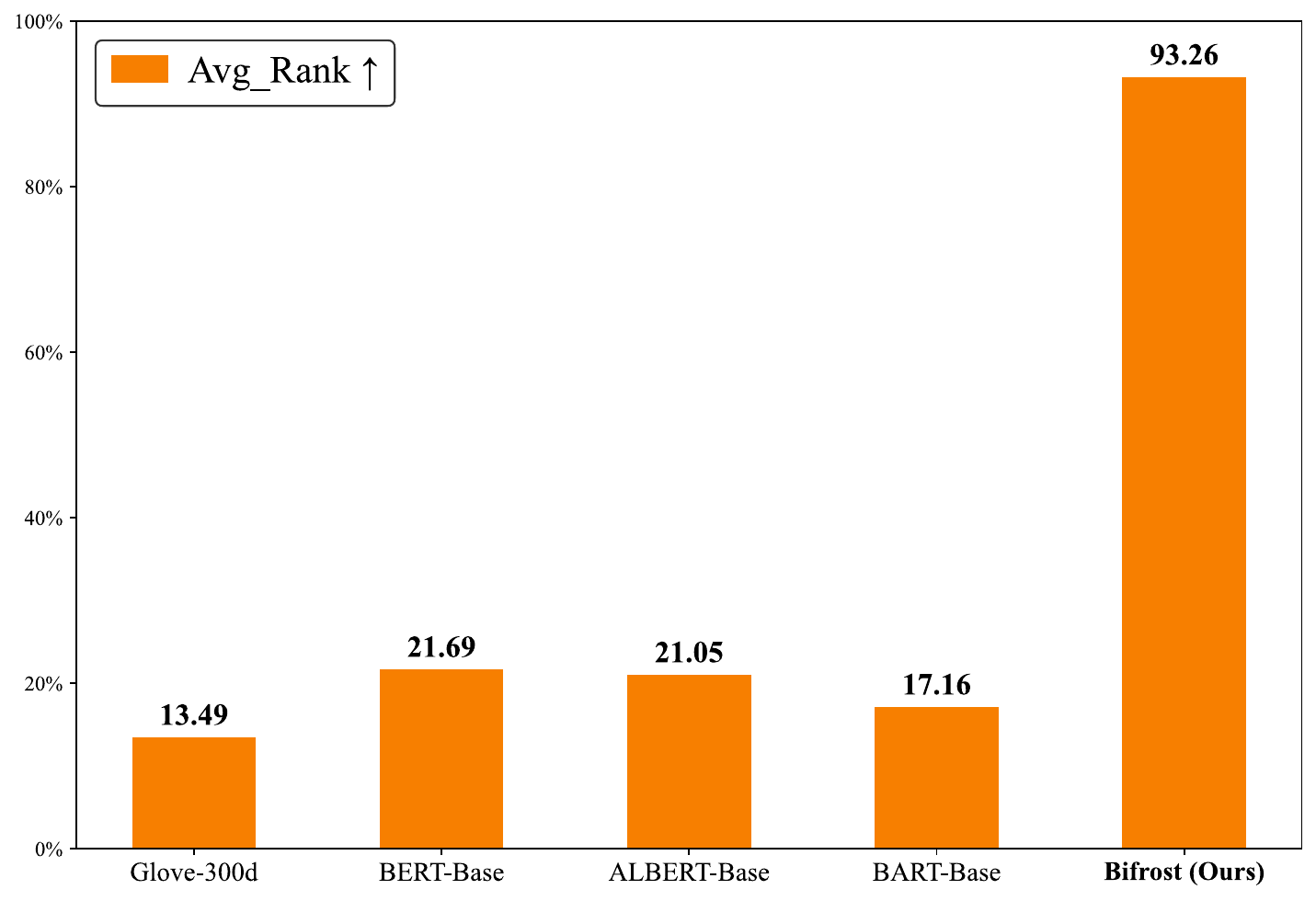}
        \caption{\footnotesize Avg\_Rank as Metric}
        \label{fig:empirical_ad_exp2}
    \end{subfigure}
    \setlength{\abovecaptionskip}{-1pt}
    \caption{\small Representational similarity in Bifrost.}
    \Description{Representational similarity in Bifrost.}
    \label{fig:empirical_ad_exp}
    \vspace{-0.4cm}
\end{figure}

\begin{figure}[htbp]
    \centering
    \begin{subfigure}[b]{0.48\linewidth}
        \centering
        \includegraphics[width=\linewidth]{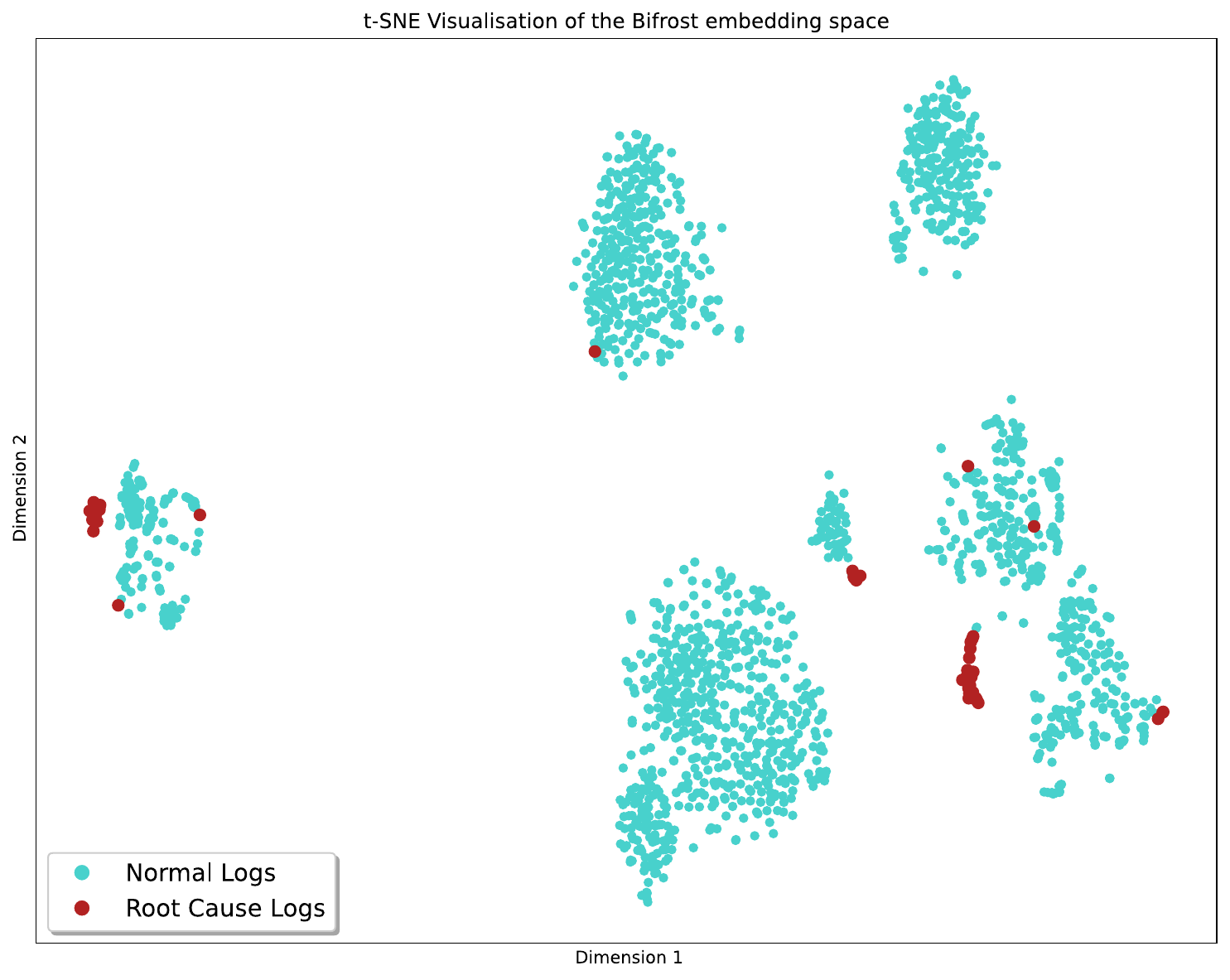}
        \caption{\footnotesize Event-level}
        \label{fig:bifrost_log}
    \end{subfigure}
    \hspace{0.02\linewidth}
    \begin{subfigure}[b]{0.48\linewidth}
        \centering
        \includegraphics[width=\linewidth]{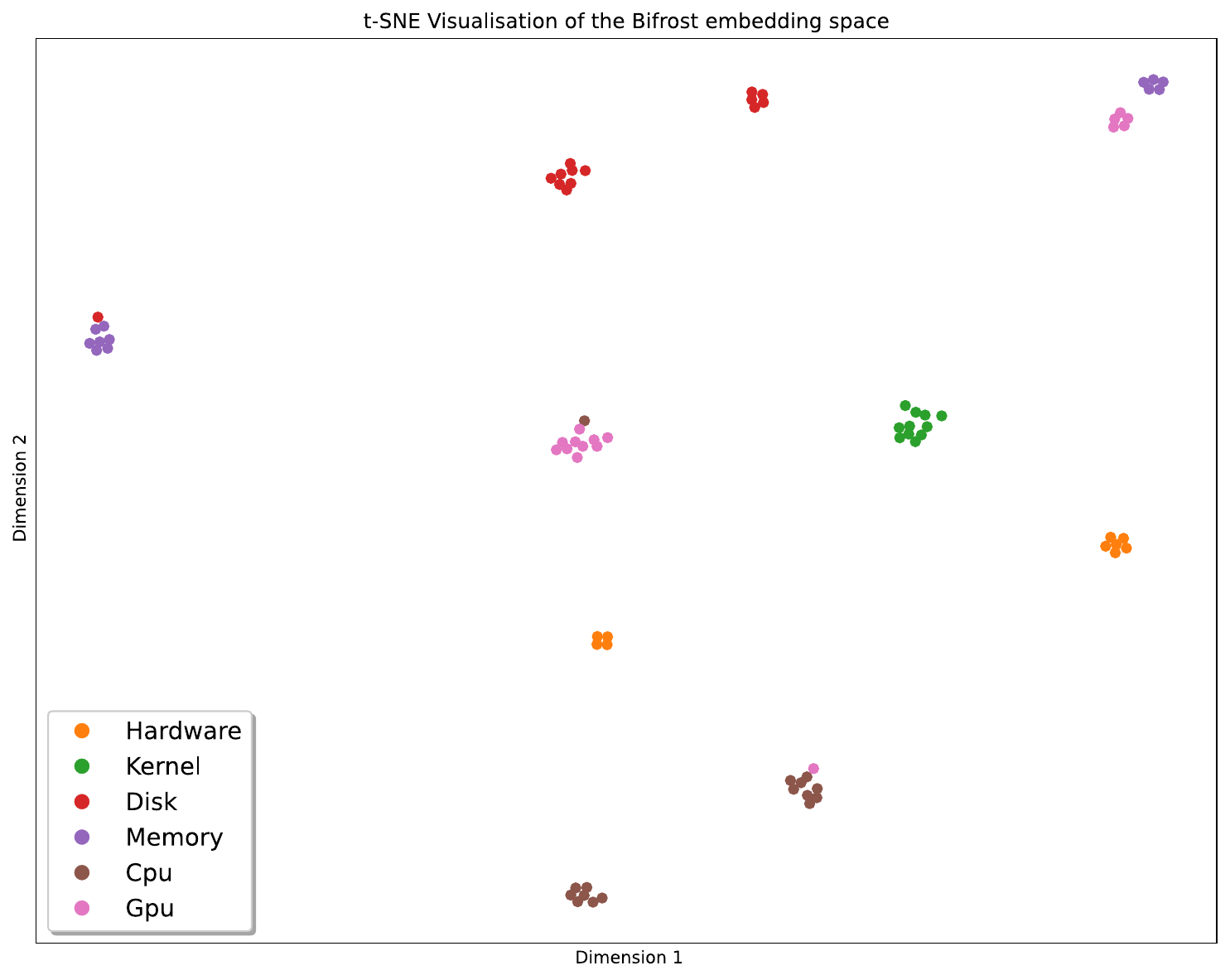}
        \caption{\footnotesize Component-level}
        \label{fig:bifrost_component}
    \end{subfigure}
    \setlength{\abovecaptionskip}{-1pt}
    \caption{\small The log representation generated by Bifrost.}
    \Description{The log representation generated by Bifrost.}
    \label{fig:bifrost_repr}
    \vspace{-0.4cm}
\end{figure}

This section evaluates whether \textsc{Bifrost} learns effective fallibility representations of logs, using the setup in Section \ref{sec:empirical}. At the execution-flow level, Figures \ref{fig:empirical_ad_exp1} and \ref{fig:empirical_ad_exp2} show that \textsc{Bifrost} performs better on both $Avg\_Sim$ and $Avg\_Rank$, indicating a stronger ability to identify unexpected execution flows. At the event level, Figure \ref{fig:bifrost_log} shows that anomalous events form more distinct clusters than those of baseline PLMs, indicating better separability. At the component level, Figure \ref{fig:bifrost_component} shows that events from the same component cluster together, reflecting stronger component awareness. Taken together, these results indicate that \textsc{Bifrost} effectively learns fallibility representations of logs through its training design, supporting effective fault diagnosis.

\section{Threats to Validity}

\textbf{Method.}
Our approach learns fallibility representations by contrastive training on injected counterfactual failures. Injection can introduce spurious cues and reduce generalizability. We mitigate this with randomized injection and self supervised training. Cross system results and ablations suggest the representations transfer beyond injected artifacts, but injection is an approximation.

\textbf{Evaluation.}
Our evaluation is constrained by dataset availability and label granularity. Accordingly, we use BGL, Thunderbird, Hadoop, and Platform-X for AD, exclude Hadoop from RCL because it lacks root-cause log labels, and evaluate FI only on Platform-X, since most existing FI datasets are built on metrics or traces rather than logs. This task-specific coverage may limit transferability, although dataset heterogeneity in AD and RCL partially mitigates this threat. FI under alternative labeling schemes remains future work.

\section{Related Work}

\subsection{Log-based Software System Fault Diagnosis}

Discovering and diagnosing software problems through log analysis has long been an active research area \cite{xia2019bugidentifier, zhang2025survey, zhang2026hypothesize, reidemeister2010identifying, huang2025no, zhang2021cloudrca}. Recent log-based fault diagnosis mainly follows two paradigms: LLM-based approaches that reason directly over raw logs or retrieved log contexts \cite{zhang2025survey, jiang2025l4, huang2025no, xiao2025coorlog, zhang2026agentic, wang2024logexpert, ma2024llmparser}, and representation-based approaches that learn log representations for downstream diagnosis. The latter mainly targets three tasks: AD, RCL, and FI. AD methods \cite{zhang2019robust, zhao2025zerolog, yang2021semi, du2017deeplog} learn normal execution patterns and detect deviations in logs. RCL methods, such as LogRCA \cite{wittkopp2024logrca} and Onion \cite{zhang2021onion}, identify logs most indicative of the root cause. FI methods \cite{duan2025famos, zhang2021cloudrca, sui2023logkg} construct deep learning models to capture the features of faults. Although they address different tasks, these methods typically rely on PLMs for log representation, which cannot bridge the structural gap between logs and natural language and thus fall short of fault diagnosis requirements. \textsc{Bifrost} addresses this by compelling PLMs to learn fallibility representations of logs through meticulously designed tasks.

\subsection{System Log Representation}
Learning log-oriented representations from system logs has been extensively studied in recent years \cite{llmelog, le2024prelog, sun2026varparser, zhao2026generality, ma2024knowlog, guo2024logformer, wu2023effectiveness}. 
LLMeLog \cite{llmelog} conducted an empirical study whose results show that log representations encoded by existing PLMs suffer from issues such as content missing, which it improves by rewriting system logs using LLMs. 
KnowLog \cite{ma2024knowlog} proposed a training framework enhanced by system knowledge, which learns the abbreviation, context, and style knowledge of logs from system logs through well-designed training tasks to improve the log representation. 
PreLog \cite{le2024prelog} is a PLM pre-trained on system logs, designed to uniformly perform various log analysis tasks through the paradigm of prompt-tuning. 
For AD tasks, LogFormer \cite{guo2024logformer} proposed a cross-system anomaly detection paradigm, which aims to learn general system representations and adapt to different systems through fine-tuning. 
Although these works all learn from system logs, they only focus on the semantic information of logs, ignoring their multi-level structural information, and are not designed for fault diagnosis tasks. 
In contrast, \textsc{Bifrost} draws inspiration from the log analysis experience of SREs, compelling PLMs to learn fallibility representations of logs to obtain high-quality log representations oriented toward fault diagnosis tasks.

\section{Conclusion}
In this paper, we propose \textsc{Bifrost}, a log representation learning method for log-based fault diagnosis. 
Inspired by the log analysis experience of SREs, \textsc{Bifrost} utilizes multi-level information from logs—such as execution flows, events, and components—as fallibility representations and learns them through meticulously designed strategies based on self-supervised contrastive learning to enhance the capabilities of PLMs in fault diagnosis. 
Comprehensive evaluations on three public systems and one industrial MLaaS system show that the log representations generated by \textsc{Bifrost} outperform existing PLMs by average margins of 9.83\% in F1 for AD, 18.28\% in HR@k for RCL, and 20.88\% in Macro-F1 for FI.
This evidence highlights the powerful role of \textsc{Bifrost} and its fallibility representations in fault diagnosis. 
In the future, we plan to explore larger-scale log representation models based on fallibility representations to achieve more effective and general fault diagnosis.

\section{Data Availability Statement}
To facilitate reproducibility, the implementation of \textsc{Bifrost} is available at \url{https://github.com/hemh02/Bifrost}. 
All public datasets used in our evaluation are sourced from LogHub~\cite{zhu2023loghub}. 
The proprietary Platform-X dataset, consisting of Company-X's production logs, cannot be publicly disclosed due to confidentiality policies. Consequently, we treat the Platform-X evaluation as an industrial case study, clearly separating its findings from claims supported by publicly available data.

\begin{acks}
This work was supported by the National Natural Science Foundation of China (Grant No. 62502015) and the Alibaba Group through Alibaba Innovative Research Program.
\end{acks}

\appendix

\section{System Components Corpus}\label{app:system-components}
In the SCP task, component semantics are represented by injecting a keyword from a predefined set of eight common system components: $\mathcal{C}=\{\texttt{cpu}, \texttt{gpu}, \texttt{disk}, \texttt{hardware}, \texttt{memory}, \texttt{kernel}, \texttt{network}, \\ \texttt{file}\}$.Because system failures typically manifest at this component granularity, and these concepts are independent of specific applications, this keyword set introduces a transferable inductive bias that significantly improves cross-system generalization.

\section{Gradient Analysis of Fallibility Jitter}
\label{sec:fj_define}\label{sec:bound_all}
We analyze fallibility representation learning from a gradient perspective and quantify the optimization error. The instability comes from the stochastic sampling in EFP and SCP, which makes each update deviate from the corresponding expected gradient. 

We reuse the notation in Section \ref{sec:method} for $\mathbf{L}$, $l_t$, $\mathbf{x}_t$, $l_i$, $\mathcal{C}$, $c_j$, $l_{t,j}=l_t \oplus c_j$, and all instances and loss terms. Let $\mathbf{W}$ denote all trainable parameters in \textsc{Bifrost}, and let $\boldsymbol{\xi}_s=(l_i,c_j)$ denote the step-wise randomness induced by context and component sampling. Eq.~\ref{eq:vanilla_loss} as $\mathcal{L}_{\text{Vanilla}}(\mathbf{W}; l_t, \boldsymbol{\xi}_s)$, define the stochastic and true gradients as $\mathbf{g}_{\text{Vanilla}}(\mathbf{W}; l_t, \boldsymbol{\xi}) = \nabla_{\mathbf{W}} \mathcal{L}_{\text{Vanilla}}(\mathbf{W}; l_t, \boldsymbol{\xi})$ and $\mathbf{G}_{\text{Vanilla}}(\mathbf{W}; l_t) = \mathbb{E}_{\boldsymbol{\xi}} \left[ \mathbf{g}_{\text{Vanilla}}(\mathbf{W}; l_t, \boldsymbol{\xi}) \right]$. Conditioned on $l_t$, their discrepancy is:
\begin{equation}
\scalebox{0.9}{
    $
    \begin{aligned}
\sigma_{\text{Vanilla}}^2(\mathbf{W}; l_t) = \mathbb{E}_{\boldsymbol{\xi}} \left[ \left\lVert \mathbf{g}_{\text{Vanilla}}(\mathbf{W}; l_t, \boldsymbol{\xi}) - \mathbf{G}_{\text{Vanilla}}(\mathbf{W}; l_t) \right\rVert^2 \right],
    \end{aligned}
    $
}
\end{equation}

Because $l_i$ and $c_j$ are sampled independently, this variance decomposes into the EFP and SCP terms:
\begin{equation}
\scalebox{0.9}{
    $
    \begin{aligned}
\sigma_{\text{Vanilla}}^2(\mathbf{W}; l_t) = \text{Var}_{l_i \in \mathbf{x}_t}\left(\mathbf{g}_{\text{EFP}}(\mathbf{W}; l_t, l_i)\right) 
+ \text{Var}_{c_j \in \mathcal{C}}\left(\mathbf{g}_{\text{SCP}}(\mathbf{W}; l_t, c_j)\right),
    \end{aligned}
    $
}
\end{equation}
where
\begin{equation}
\mathbf{g}_{\text{EFP}}(\mathbf{W}; l_t, l_i) = \nabla_{\mathbf{W}}(\lambda_1 \mathcal{L}_{\text{EFP}}),
\end{equation}
\begin{equation}
\mathbf{g}_{\text{SCP}}(\mathbf{W}; l_t, c_j) = \nabla_{\mathbf{W}}(\lambda_3 \mathcal{L}_{\text{SCP}}).
\end{equation}
Under this conditioning, the AED term does not contribute because it does not depend on $\boldsymbol{\xi}$.

Let $s$ be the training step, $\eta_s$ the learning rate, and $\Delta(\cdot, \cdot)$ the optimizer update direction. The ideal and stochastic trajectories are \scalebox{0.9}{$\mathbf{W}_{s+1} = \mathbf{W}_s - \eta_s \Delta(\mathbf{W}_s, \mathbf{G}_s)$} and \scalebox{0.9}{$\mathbf{W}'_{s+1} = \mathbf{W}'_s - \eta_s \Delta(\mathbf{W}'_s, \mathbf{g}_s)$}, where \\ \scalebox{0.9}{$\mathbf{G}_s(\mathbf{W}) \triangleq \mathbf{G}(\mathbf{W}; l_s)$} and \scalebox{0.9}{$\mathbf{g}_s(\mathbf{W}) \triangleq \mathbf{g}(\mathbf{W}; l_s, \boldsymbol{\xi}_s)$}. We define the resulting trajectory mismatch as \scalebox{0.9}{$\mathbf{E}_s = \mathbf{W}_s - \mathbf{W}'_s$}, which is the fallibility jitter.

\textbf{Weight Error Upper Bound.}
For standard SGD, the update vector is $\Delta(\mathbf{W}, \mathbf{g}) \triangleq \mathbf{g}$. The single-step evolution of the weight error is:
\begin{equation}
\scalebox{0.9}{
    $
    \begin{aligned}
\mathbf{E}_{s+1} &= \mathbf{W}_{s+1} - \mathbf{W}'_{s+1}
\\
&= (\mathbf{W}_s - \eta_s \mathbf{G}_s(\mathbf{W}_s)) - (\mathbf{W}'_s - \eta_s \mathbf{g}_s(\mathbf{W}'_s)) \\
&= \mathbf{E}_s - \eta_s (\mathbf{G}_s(\mathbf{W}_s) - \mathbf{g}_s(\mathbf{W}'_s)),
    \end{aligned}
    $
}
\end{equation}
Applying the triangle inequality and the Lipschitz assumption on $\mathbf{G}_s(\cdot)$, and defining $\boldsymbol{\delta}_s(\mathbf{W})=\mathbf{g}_s(\mathbf{W})-\mathbf{G}_s(\mathbf{W})$, we obtain
\begin{equation}
\scalebox{0.9}{
    $
    \begin{aligned}
    \left\lVert \mathbf{G}_s(\mathbf{W}_a) - \mathbf{G}_s(\mathbf{W}_b) \right\rVert
    \le L_{\text{Lip}} \left\lVert \mathbf{W}_a - \mathbf{W}_b \right\rVert,
    \end{aligned}
    $
}
\end{equation}

\begin{equation}
\scalebox{0.9}{
    $
    \begin{aligned}
    \left\lVert \mathbf{E}_{s+1} \right\rVert
    \le (1 + \eta_s L_{\text{Lip}}) \left\lVert \mathbf{E}_s \right\rVert
    + \eta_s \left\lVert \boldsymbol{\delta}_s(\mathbf{W}'_s) \right\rVert,
    \end{aligned}
    $
}
\end{equation}
Assuming a constant learning rate $\eta_s = \eta$, unrolling the recursion, taking expectation over $\{\boldsymbol{\xi}_s\}_{s=0}^{S-1}$, and using $\mathbb{E}_{\boldsymbol{\xi}_s}[\left\lVert \boldsymbol{\delta}_s(\mathbf{W}'_s) \right\rVert^2 \vert \mathbf{W}'_s] = \sigma^2(\mathbf{W}'_s; l_s)$ yield the final upper bound:
\begin{equation}
\mathbb{E}[\left\lVert \mathbf{E}_S \right\rVert]
\le \eta \sum_{s=0}^{S-1} (1+\eta L_{\text{Lip}})^{S-1-s}\, \mathbb{E}_{\mathbf{W}'_s}[\sigma(\mathbf{W}'_s; l_s)],
\end{equation}
This bound shows how per-step sampling variance accumulates into weight error, linking stochasticity to representation instability.

\section{Weight Error Upper Bound Analysis for CARL}
\label{sec:bound_carl}
Given the upper bound in Sec.~\ref{sec:bound_all}, it suffices to show that CARL reduces the per-step gradient variance relative to the vanilla strategy. We show that this happens through two mechanisms: variance isolation for the context-sampling randomness and a negative covariance constraint for the component-sampling randomness.

\textbf{Variance Isolation.}
In CARL, the randomness introduced by context sampling $l_i$ is isolated to an independent term, $\mathcal{L}_{\text{CL}}$, with weight $\lambda_2$, whereas in the vanilla strategy the entire EFP term with weight $\lambda_1$ depends on $l_i$. Assuming comparable gradient scales in the two strategies and $\lambda_2 < \lambda_1$, the variance contribution from $l_i$ is reduced:
\begin{equation}
\scalebox{1}{
    $
\begin{aligned}
\text{Var}_{l_i}(\nabla_{\mathbf{W}}(\lambda_2 \mathcal{L}_{\text{CL}})) 
&\propto \lambda_2^2 \cdot \text{Var}_{l_i}(\nabla_{\mathbf{W}}\mathcal{L}_{\text{CL}}) \\
&< \lambda_1^2 \cdot \text{Var}_{l_i}(\nabla_{\mathbf{W}}\mathcal{L}_{\text{EFP}}),
\end{aligned}
    $
}
\end{equation}

\textbf{Negative Covariance Constraint.}
CARL also reduces the variance introduced by component sampling. When an anchor $l_{t,j}=l_t \oplus c_j$ is sampled for the SCP task, the remaining component-injected logs $\{l_{t,i}\}_{i \neq j}$ are reused as shared negatives for other tasks. For a sampled component $c_j$, the corresponding gradient decomposes as $\mathbf{G}(c_j)=\mathbf{A}(c_j)+\mathbf{B}(c_j)$, where
\begin{equation} \mathbf{A}(c_j) \triangleq \nabla_{\mathbf{W}}(\lambda_4 \mathcal{L}_{\text{SCP}}(\text{anchor}=l_{t,j}, \dots)), \end{equation}
\begin{equation}
    \scalebox{0.9}{
        $
    \begin{aligned}
    \mathbf{B}(c_j) \triangleq \nabla_{\mathbf{W}} \sum_{\text{task} \in \{\text{EFP}, \text{CL}, \text{AED}\}} \lambda_{\text{task}} \mathcal{L}_{\text{task}}(\dots, \text{Negatives}=\{l_{t,i}\}_{i\neq j}),
    \end{aligned}
        $
    }
\end{equation}
and therefore
\begin{equation}
\scalebox{1}{
    $
\begin{aligned}
\text{Var}(\mathbf{G}) = \text{Var}(\mathbf{A}) + \text{Var}(\mathbf{B}) + 2\text{Cov}(\mathbf{A}, \mathbf{B}),
\end{aligned}
    $
}
\end{equation}

Let $\mathbf{A}(c_i) \triangleq \mathbf{g}_{A,i}$ and $\mathbf{B}(c_j) \triangleq \sum_{i \neq j} \mathbf{g}_{B,i}$. Since both $\mathbf{g}_{A,i}$ and $\mathbf{g}_{B,i}$ push away $l_{t,i}$, they are positively correlated:
\begin{equation}
\scalebox{1}{
    $
\begin{aligned}
\text{Cov}_i(\mathbf{g}_{A,i}, \mathbf{g}_{B,i}) = \mathbb{E}_i[\mathbf{g}_{A,i}^T \mathbf{g}_{B,i}] - \mathbb{E}_i[\mathbf{g}_{A,i}]^T \mathbb{E}_i[\mathbf{g}_{B,i}] > 0,
\end{aligned}
    $
}
\end{equation}

Let $\lvert\mathcal{C}\rvert=N$, assume uniform sampling, and define 
\begin{equation}
\bar{\mathbf{g}}_A=\frac{1}{N}\sum_{i=1}^N \mathbf{g}_{A,i}
\end{equation}
\begin{equation}
\bar{\mathbf{g}}_B=\frac{1}{N}\sum_{i=1}^N \mathbf{g}_{B,i}
\end{equation}
Because the sampled anchor index $j$ is excluded from the shared-negative term $\mathbf{B}(c_j)$, a direct expansion gives
\begin{equation}
\scalebox{1.0}{
    $
\begin{aligned}
\text{Cov}(\mathbf{A}, \mathbf{B}) &= \left( N \bar{\mathbf{g}}_A^T \bar{\mathbf{g}}_B - \mathbb{E}_j[\mathbf{g}_{A,j}^T \mathbf{g}_{B,j}] \right) - (\bar{\mathbf{g}}_A^T) ((N-1)\bar{\mathbf{g}}_B) \\
&= \bar{\mathbf{g}}_A^T \bar{\mathbf{g}}_B - \mathbb{E}_j[\mathbf{g}_{A,j}^T \mathbf{g}_{B,j}] \\
&= - \left( \mathbb{E}_j[\mathbf{g}_{A,j}^T \mathbf{g}_{B,j}] - \bar{\mathbf{g}}_A^T \bar{\mathbf{g}}_B \right) \\
&= - \text{Cov}_j(\mathbf{g}_{A,j}, \mathbf{g}_{B,j}),
\end{aligned}
    $
}
\end{equation}

Since $\text{Cov}_j(\mathbf{g}_{A,j}, \mathbf{g}_{B,j}) > 0$, we have $\text{Cov}(\mathbf{A}, \mathbf{B}) < 0$, so the shared-negative design cancels part of the variance caused by component sampling.

Taken together, CARL reduces both the $l_i$-induced and $c_j$-induced gradient variance. By Sec.~\ref{sec:bound_all}, this yields a tighter upper bound on the cumulative weight error and mitigates fallibility jitter.

\balance 
\bibliographystyle{ACM-Reference-Format}
\bibliography{sample-base}

\end{document}